\documentclass[aps,prd,amsmath,amssymb,twocolumn,superscriptaddress,preprintnumbers,nofootinbib]{revtex4-1}
\pdfoutput=1
\usepackage{graphicx}
\usepackage{amsthm}
\usepackage{amsmath}
\usepackage{graphicx,subfigure}
\usepackage{dcolumn}
\usepackage{bm}
\usepackage{slashed}

\newcommand{\be}{\begin{eqnarray}}
\newcommand{\ee}{\end{eqnarray}}
\newcommand{\bea}{\begin{eqnarray}}
\newcommand{\eea}{\end{eqnarray}}

\newcommand{\keV}{{~\rm keV}}
\newcommand{\MeV}{{~\rm MeV}}
\newcommand{\GeV}{{~\rm GeV}}
\newcommand{\gev}{{~\rm GeV}}
\newcommand{\TeV}{{~\rm TeV}}
\newcommand{\tev}{{~\rm TeV}}

\newcommand{\vev}[1]{{\langle #1 \rangle}}

\newcommand{\mX}{m_{_\chi} }
\newcommand{\mXp}{m_{_{\chi^*}} }
\newcommand{\thetaX}{\theta_{_\chi}}
\newcommand{\thetaXS}{\theta_{_\chi}}
\newcommand{\thetaXA}{\theta_{_\chi}}
\newcommand{\mN}{m_{_N}}
\newcommand{\mZ}{m_{\rm Z}}
\newcommand{\mW}{m_{\rm W}}
\newcommand{\LamR}{\Lambda_{_R}}

\newcommand{\muN}{\mu_{_N}}
\newcommand{\muther}{\mu_{\rm thermal}}
\newcommand{\mug}{\mu_{_{\gamma}}}
\newcommand{\muZ}{\mu_{_{\rm Z}}}

\newcommand{\RN}{R_{_N}}

\newcommand{\ZZ}{{\rm Z}}
\newcommand{\Wpm}{\rm W^\pm}
\newcommand{\DeltaM}{\Delta M}
\newcommand{\cW}{{\rm c_{_W}}}
\newcommand{\sW}{{\rm s_{_W}}}
\newcommand{\cX}{{\rm c_{_\chi}}}
\newcommand{\sX}{{\rm s_{_\chi}}}
\newcommand{\SUWeak}{{\rm SU_{_W}(2)}}
\newcommand{\phiS}{s}
\newcommand{\phiA}{a}

\begin{document}

\title{How Dark Are Majorana WIMPs? Signals from MiDM and Rayleigh Dark Matter}
\author{Neal Weiner}
\email{neal.weiner@nyu.edu}
\affiliation{Center for Cosmology and Particle Physics, Department of Physics, New York University, New York, NY 10003}
\author{Itay Yavin}
\email{iyavin@perimeterinstitute.ca}
\affiliation{Department of Physics \& Astronomy, McMaster University 1280 Main St. W. Hamilton, Ontario, Canada, L8S 4L8}
\affiliation{Perimeter Institute for Theoretical Physics 31 Caroline St. N, Waterloo, Ontario, Canada N2L 2Y5.}


\begin{abstract}
The effective interactions of dark matter with photons are fairly restricted. Yet both direct detection as well as monochromatic $\gamma$ ray signatures depend sensitively on the presence of such interactions. For a Dirac fermion, electromagnetic dipoles are possible, but are very constrained. For Majorana fermions, no such terms are allowed. We consider signals of an effective theory with a Majorana dark matter particle and its couplings to photons. In the presence of a nearby excited state, there is the possibility of a magnetic dipole transition (Magnetic inelastic Dark Matter or MiDM), which yields both direct and indirect detection signals, and, intriguingly, yields essentially the same size over a wide range of dipole strengths. Absent an excited state, the leading interaction of WIMPs is similar to the Rayleigh scattering of low energy photons from neutral atoms, which may be captured by an effective operator of dimension 7 of the form $\bar{\chi}\chi F_{\mu\nu}F^{\mu\nu}$. While it can be thought of as a phase of the Magnetic inelastic Dark Matter scenario where the excited state is much heavier than the ground state, it can arise from other theories as well. We study the resulting phenomenology of this scenario: gamma ray lines from the annihilation of WIMPs; nuclear recoils in direct detection; and direct production of the WIMP pair in high-energy colliders. Considering recent evidence in particular for a 130 GeV line from the galactic center, we discuss the detection prospects at upcoming experiments.
\end{abstract}

\maketitle

\section{Introduction}

Weakly interacting massive particles (WIMPs) have long been studied as potential candidates for the cold dark matter observed in the Universe. The most well-motivated and deservedly most well-studied WIMPs are those that emerge in extensions of the Standard Model associated with the seemingly unrelated problems of the electroweak scale, such as supersymmetric extensions. An orthogonal line of inquiry is motivated by the deceptively elementary question of ``how dark is Dark Matter?'' Namely, what are the strongest constraints on the interaction of dark matter with the electromagnetic field? Numerous studies already exist and in particular the idea of electric and magnetic dipole interactions have recently attracted considerable attention~\cite{Bagnasco:1993st,Pospelov:2000bq,Sigurdson:2004zp,Gardner:2008yn, Masso:2009mu,Cho:2010br,An:2010kc,McDermott:2010pa,Chang:2010en,Banks:2010eh,DelNobile:2012tx}. In these models, single photon exchange provides a possible direct detection signal, while annihilation into two photons might provide an indirect detection signal (see e.g., \cite{Goodman:2010qn}). 

However, if dark matter is a Majorana fermion, then these single-photon couplings through electromagnetic dipoles do not exist - the dipole operator vanishes identically for Majorana fermions. For Dirac fermions, it is naturally off-diagonal~\cite{Dreiner:2008tw,Kopp:2009qt}. For pseudo-Dirac fermions, in which case the ground state is the dark matter candidate, the dipole interaction mediates transitions between this ground state $\chi$ and an excited state $\chi^*$. The authors of Ref.~\cite{Chang:2010en} exploited this possibility to build a model, dubbed Magnetic inelastic Dark Matter (MiDM), to explain the DAMA results through dipole-dipole dominated scattering. The interaction Lagrangian of MiDM is
\be
\label{eqn:MiDMInteraction}
\mathcal{L}= \left(\frac{\mu_\chi}{2}\right)\bar \chi^* \sigma_{\mu\nu} B^{\mu\nu} \chi + c.c.,
\ee 
where $\mu_\chi$ is the dipole strength, $B^{\mu\nu}$ is the hypercharge field-strength tensor, and $\sigma_{\mu\nu} = i[\gamma_\mu,\gamma_\nu]/2$. This coupling contains within it the interaction with the electromagnetic field. We study the signatures of this model in the first part of this paper. 

In the limit that we take the excited state heavy, we are left with a Majorana fermion, and we can again ask the question ``how dark is Dark Matter?''. Starting with the MiDM Lagrangian above if the excited state, $\chi^*$, is much heavier than the energy available then it can be integrated out to yield the interactions 
\be
\label{eqn:RDMInteraction}
\mathcal{L} = \frac{\mu_\chi^2}{2\mXp} \left(\bar{\chi}\chi B_{\mu\nu}B^{\mu\nu} + i \bar{\chi}\gamma_5\chi B_{\mu\nu}\tilde{B}^{\mu\nu}\right),
\ee 
where $ \tilde{B}^{\mu\nu} = \tfrac{1}{2}\epsilon^{\mu\nu\alpha\beta}B_{\alpha\beta}$ and $\epsilon^{\mu\nu\alpha\beta}$ is the Levi-Civita symbol. Motivated by this form, in the second part of this paper we will concentrate on the slightly more general case for the interaction of DM with the electroweak field strengths 
\be
\label{eqn:RDMLagrangian}
\mathcal{L} = &\frac{1}{4\LamR^3}&~\Big\{ \bar{\chi}\chi \left( \cos\thetaXS B_{\mu\nu}B^{\mu\nu} + \sin\thetaXS{\rm Tr} W_{\mu\nu}W^{\mu\nu} \right) \\\nonumber  &+&\left. i~\bar{\chi}\gamma_5\chi \left( \cos\thetaXA B_{\mu\nu}\tilde{B}^{\mu\nu} + \sin\thetaXA{\rm Tr} W_{\mu\nu}\tilde{W}^{\mu\nu} \right)\right\}.
\ee 
Here $\thetaX$ quantifies the relative coupling to the field strength of hypercharge in comparison to that of $\SUWeak$ and $\LamR$ is some high scale related to the cut-off scale of the  theory. We will discuss UV realizations in a later section, but simple scenarios can arise either as a limit of MiDM, or for instance integrating out a dilaton (or axi-dilaton). The interactions of Eq.~(\ref{eqn:RDMLagrangian}) are akin to the familiar interactions of photons with neutral atoms at long wavelengths that lead to Rayleigh scattering. Hence we dub this scenario \textsl{Rayleigh Dark Matter} (RayDM). This could be the entirety of the DM interaction with the standard model, but it also serves as a reasonable form of the effective operators responsible for $\gamma$ lines in many models (even when they freeze out dominantly through other channels). The special form of this interaction, which necessitates at least two force mediators, requires a reconsideration of the basic processes by which we hope to detect dark matter and this constitutes a part of the current work. 

In this paper we set to explore these different possibilities for the interaction of Majorana WIMPS with light. The paper is organized as follows: In section~\ref{sec:MiDM} we discuss in detail the MiDM scenario including its signatures in gamma rays as well as the prospects for seeing it in direct detection experiments;   In section~\ref{sec:RayDM} we explore the phenomenology of RayDM;  Section~\ref{sec:collider} is devoted to the prospects of collider searches for both MiDM as well as RayDM; Finally, the main findings of this work are summarized in the conclusions, section~\ref{sec:conclusions}. We caution the reader that the clear separation between MiDM and RayDM is not always appropriate. As we shall discuss and emphasize below, there are certain aspects of the phenomenology where the two scenarios and the operators involved cannot be logically separated.   

\section{$\text{MiDM}$}
\label{sec:MiDM}

In this section we concentrate on the MiDM scenario, but consider a slightly more general form of the magnetic dipole  interactions
\be
\label{eqn:generalMiDMInteraction}
\mathcal{L}= \left(\frac{\mug}{2}\right)\bar \chi^* \sigma_{\mu\nu} F^{\mu\nu} \chi + \left(\frac{\muZ}{2}\right)\bar \chi^* \sigma_{\mu\nu} Z^{\mu\nu} \chi + c.c., 
\ee 
where $F_{\mu\nu}$ and $Z_{\mu\nu}$ are respectively the field strength of the photon and $\ZZ$ boson, $\mug$ and $\muZ$ are the corresponding dipole strength, and $\chi$ and $\chi^*$ are two Weyl fermions. If the interaction above the electroweak scale is entirely with the field strength of hypercharge then $\muZ/\mug = - \tan\theta_{_W}$. In what follows we explore the more general possibility since one can entertain additional operators with the field strength of $\SUWeak$ that would result in a different ratio. An example of such an operator is the dimension 7 operator $\left(\frac{\mug}{2}\right)\bar \chi^* \sigma_{\mu\nu} \chi {\rm Tr} ~ h^{\dag} W^{\mu\nu} h$. However, considering that such operators are generically further suppressed we expect the deviation away from the relation $\muZ/\mug = -\tan\theta_{_W}$ to be small. 

The phenomenology of this theory depends crucially on the mass splitting between the two states, $\chi$ and $\chi^*$. When the mass splitting is large $\mXp - \mX \gtrsim \mX$ the phenomenology is similar to that of RayDM, which is described in the next section. More generally, even for a smaller splitting such that $\mXp \gtrsim \mX/20$ the heavier state $\chi^*$ is not present during the early universe freeze-out of the lighter state $\chi$. In this case, unless other channels are available through new interactions, the relic abundance is determined by the annihilation into photons and vector-bosons, which typically requires larger dipole strength.  The prospects for direct detection in this case are fairly gloomy, but collider constraints provide strong and interesting bounds on this possibility as we describe in section~\ref{sec:collider}.

As a consequence, in this section we focus on the case when the mass splitting is small. In particular, when the mass splitting to the excited state vanishes or is smaller than the kinetic energy of the WIMP in the halo $\Delta M = \mXp - \mX  \sim 100\keV$ the cross-section of scattering on the nucleus is much larger and the corresponding rates in direct detection experiments are phenomenologically interesting. 

Before moving onto the details, it is important to make the phenomenological point: {\em for thermally produced MiDM, the direct and gamma-ray line indirect signatures are roughly independent of the size of the dipole}\footnote{Since the breaking of hypercharge makes the phenomenology depends in principle on both $\mug$ as well as $\muZ$, this statement assumes the absence of any unexpectedly large difference in scale between $\muZ$ and $\mug$.}. That is to say, even if MiDM is a sub-dominant component of the dark matter, the amplitude of these signals would be unchanged.

This arises quite simply. The annihilation rate is governed by the annihilation of WIMPs into fermion pairs~\cite{Fortin:2011hv} which scales as $\mu_\chi^2$. This implies that the number density of WIMPs scales as $n_{\chi} \sim \mu_\chi^{-2}$, i.e.,
\be
\rho_{\rm MiDM} = \rho_0 \times \frac{\muther^2}{\mu_\chi^2},
\ee
where $\muther$ is the dipole needed to achieve the appropriate relic abundance that arises from a cross section of $6\times 10^{-26} {\rm cm}^3 {\rm s}^{-1}$. For a WIMP of mass $\mX = 130\GeV$ we find $\muther \approx 1.2 \times 10^{-3} \muN$ ($2.0 \times 10^{-3} \muN$) for a dipole ratio of $\muZ/\mug = -\tan\theta_{_W}$ ($\muZ/\mug = 0$). Here $\muN = 0.16\GeV^{-1}$ is the nuclear magneton. This dipole strength corresponds to an annihilation rate into gamma rays of $\sigma(\chi\chi\rightarrow \gamma\gamma) = 2.5\times 10^{-29}~{\rm cm^3/s}$ ($ 1.6\times 10^{-28}~{\rm cm^3/s}$ ). 

The collision rate in direct detection experiments scales as $n_\chi \sigma_{\chi N}$, thus
\be
R_{DD} \propto n_\chi \mu_\chi^2 = \frac{\rho_0}{m_\chi}\frac{\muther^2}{\mu_\chi^2} \times \mu_\chi^2 = \frac{\rho_0}{m_\chi} \muther^2,
\ee
where $\rho_0 \approx 0.4 \gev/{\rm cm}^3$ is the local density of a WIMP comprising all of the dark matter. Such a scaling phenomenon is well known in many WIMP models, that when s-channel annihilation diagrams are directly linked to t-channel scattering, the lower relic abundance is compensated by the higher scattering cross section (see, e.g., \cite{Duda:2001ae})\footnote{We do not consider the implications of a CP violating inelastic electric dipole moment here. Due to a velocity-unsuppressed dipole-charge scattering (see e.g., \cite{Pospelov:2000bq}), the constraints in \cite{Fortin:2011hv} constrain the dipole $\mu_\chi \lesssim 10^{-8} \muN$. At these levels the indirect signals would be negligible, unless the excited state is inaccessible.}. 

A possibly more remarkable scaling is associated with the $\gamma \gamma$ and $\gamma\ZZ$ signatures. The cross section for these processes is proportional to $\mu_\chi^4$. Thus, the indirect $\gamma$-ray rate scales as
\be
R_{\gamma \gamma} \propto n_\chi^2 \mu_\chi^4 =  \frac{\rho_0^2}{m_\chi^2}\frac{\muther^4}{\mu_\chi^4} \times \mu_\chi^4 = \frac{\rho_0^2}{m_\chi^2} \muther^4.
\label{eq:scalingindirect}
\ee
Again, if the WIMP is thermal, the $\gamma \gamma$ signal is {\em independent} of the size of the dipole, even if the fraction of the dark matter is much smaller. There are important caveats to this, as we shall discuss below, but they do not change the fact that even for very large dipoles (yielding under-abundant dark matter) the signals are at the same level as that of a thermal dominant WIMP\footnote{Once $\mu_\chi$ is large enough this scaling ceases because the annihilation to gauge bosons dominates. For 130 GeV this occurs at $\mu_\chi\gtrsim .05 \mu_N$. At this size, we shall see that collider constraints would exclude the scenario already.}.

However, such a scenario is excluded unless $\delta = m_{\chi^*}-m_\chi \gtrsim 1/2 m_\chi v^2$ and the direct detection scattering is either inelastic (if the excited state is accessible) or not present (if it is not). Intriguingly, the size of the signals would be appropriate for DAMA~\cite{Bernabei:2010mq} (via the MiDM scenario~\cite{Chang:2010en}) and approximately for the recently reported 130 GeV signal, given the astrophysical uncertainties~\cite{Bringmann:2012vr, Weniger:2012tx,Tempel:2012ey,Su:2012ft}.

\subsection{Annihilation Rate and Relic Abundance}
To understand these points in detail, let us consider the precise values realized. 
When the excited state $\chi^*$ is present, the annihilation into Standard Model fermion pairs through $\gamma/\ZZ$ dominates. The annihilation cross-section to leading order in velocity is given by
\be
\label{eqn:MiDMannffbar}
\sigma(\chi\chi^*\rightarrow f\bar{f}) v = \alpha q_f^2 \mug^2  \Big(&1& +~ 2 v_f \frac{\muZ}{\mug}~ \xi( 4\mX^2 ) \\ \nonumber &+& (v_f^2 + a_f^2)  \frac{\muZ^2}{\mug^2}  ~\xi^2( 4\mX^2) \Big). 
\ee
Here $q_f$ is the fermion's electric charge, $v_f$ ($a_f$) is the ratio of its vector (axial-vector) coupling to the Z boson to its electromagnetic coupling, and $\xi(s) = s/(s-\mZ^2)$. In the above expression we took $\mXp = \mX$ since it is only when the mass splitting is not too large that the heavier state is relevant. The annihilation rates of the lighter state into vector-bosons are
\be
\label{eqn:MiDManngg}
\sigma(\chi\chi^*\rightarrow \gamma\gamma) v &=& \frac{\mX^4}{4\pi} \left(\frac{2\mug^2}{\mXp}\right)^2\frac{1}{\left(1+\mX^2/\mXp^2\right)^2}, \\ 
\label{eqn:MiDManngZ}
\sigma(\chi\chi^*\rightarrow \gamma\ZZ) v &=&2\frac{\mX^4}{4\pi} \left(\frac{2\mug^2}{\mXp}\frac{2\muZ^2}{\mXp}\right) \left(1-\frac{\mZ^2}{4\mX^2} \right)^{3} \\ \nonumber &\times&\left(1+\frac{\mZ^2}{4\mX\mXp} \right)^2/\left(1+\frac{2\mX^2-\mZ^2}{2\mXp^2}\right)^2, \\
%
\label{eqn:MiDMannZZ}
\sigma(\chi\chi^*\rightarrow \ZZ\ZZ) v &=& \frac{\mX^4}{4\pi} \left(\frac{2\muZ^2}{\mXp}\right)^2 \left(1-\frac{\mZ^2}{\mX^2} \right)^{3/2} \\ \nonumber &\times&  \left(1+\frac{\mZ^2}{2\mX\mXp}\right)^2 /\left(1+ \frac{\mX^2-\mZ^2}{\mXp^2} \right)^2.
\ee
In the case of large mass splittings, we are effectively left with a single species at freeze-out and its annihilation into vector-bosons must therefore be sufficiently large so as to yield $\sigma_{\rm tot} v \approx 3 \times 10^{-26}{\rm cm^3/s}$. This implies a rather large dipole strength which is in tension with collider searches for mono-photons discussed in section~\ref{sec:collider}. 

In the case of intermediate and small mass splittings, the cosmological history is slightly different and the necessary total annihilation rate is consequently altered. Since the heavier state decays to the lighter state through the dipole transition only after freeze-out, we effectively have two species during freeze-out, each of which can only annihilate on the other. The relic abundance necessary at freeze-out is therefore only half its usual value\footnote{This is true only as long as the heavier state's lifetime is sufficiently short so that present day DM is entirely composed of the lighter state. For $\mXp-\mX \gtrsim 100\keV$ the lifetime is shorter than about a microsecond.}. This implies that the total annihilation rate is larger $\sigma_{\rm tot} v \approx 6 \times 10^{-26}{\rm cm^3/s}$. Interestingly, since this requires a dipole strength larger by a factor of $\sqrt{2}$, this leads to an increase of the annihilation rate into $\gamma\gamma$ and $\gamma\ZZ$ of factor by 4. 

\begin{figure}
\begin{center}
\includegraphics[width=0.45 \textwidth]{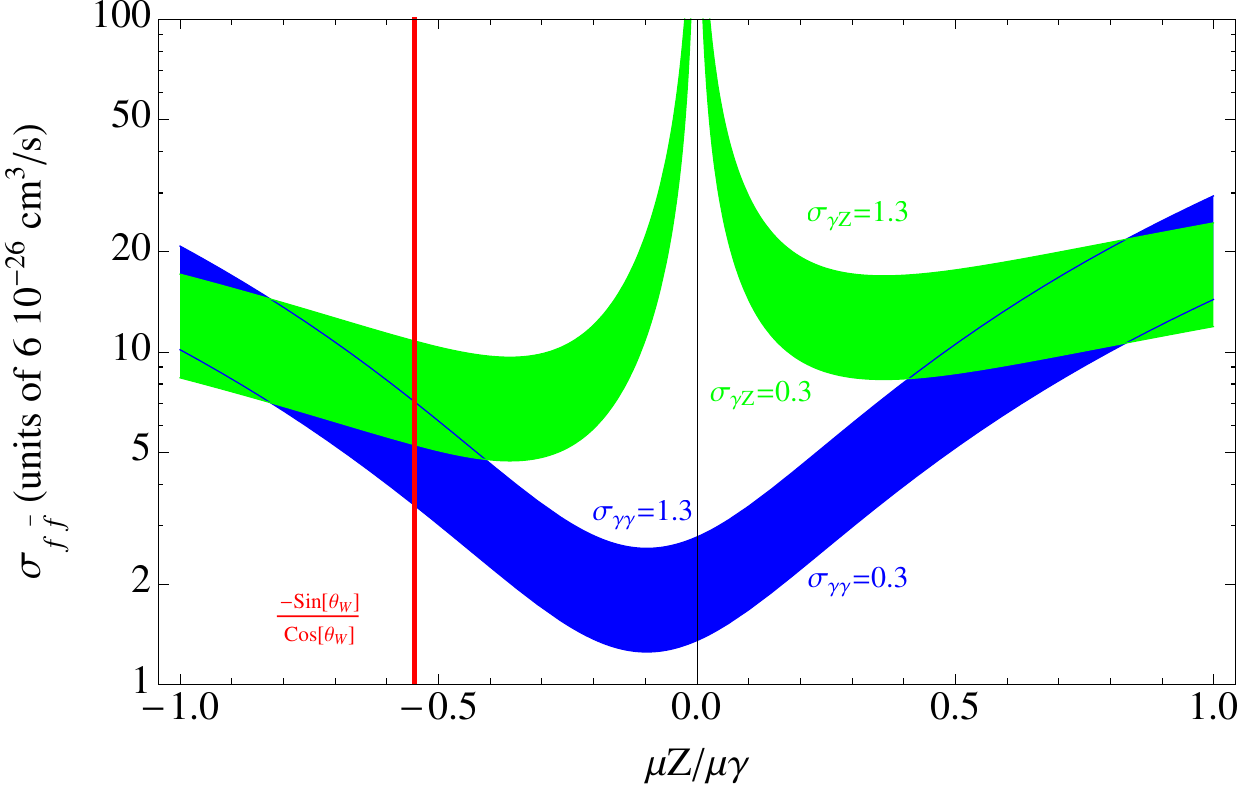}
\end{center}
\caption{The total annihilation rate for $\mX =\mXp = 130\GeV$ to fermion pairs, $\sigma(\chi\chi\rightarrow \gamma\gamma)$ in the MiDM scenario as a function of the relative strength of the photon dipole to $\ZZ$ dipole. The dipole itself is normalized to yield the annihilation rate into $\gamma\gamma$ (blue band) and $\gamma\ZZ$ (green band). The bands span the range shown in units of $10^{-27}{\rm cm^3/s}$. The red vertical line shows the relevant ratio of dipoles in the case of MiDM interaction with hypercharge only.  }
\label{fig:MiDM_annihilation}
\end{figure}

In two recent papers, Refs.~\cite{Bringmann:2012vr, Weniger:2012tx} reported on a tentative gamma ray line in the Fermi/LAT data at $E_{\gamma} = 130\GeV$, which has recently been confirmed by Tempel et al.~\cite{Tempel:2012ey} and Su and Finkbeiner \cite{Su:2012ft}. This result can be accommodated within the MiDM scenario with a WIMP mass $\mX = 130\GeV$ leading to the gamma ray line at $E_{\gamma} = 130\GeV$ through $\chi\chi\rightarrow \gamma\gamma$. It will generically also result in an additional line of comparable strength at $E_{\gamma} = \mX-\mZ^2/4\mX = 114\GeV$ from $\chi\chi\rightarrow \gamma{\rm Z}$, which is consistent with the data (see also the discussion in ref.~\cite{Rajaraman:2012db})\footnote{Alternatively, $\mX = 144\GeV$ may give rise to the line at $E_{\gamma} = 130\GeV$ through the annihilation $\chi\chi\rightarrow \gamma{\rm Z}$. In this case the annihilation to two photons would have to be somewhat suppressed since no feature is observed at $E_{\gamma} = 144\GeV$. Such a suppression is more natural in the RayDM scenario discussed in the next section.}. In Fig.~\ref{fig:MiDM_annihilation} we plot the annihilation rate into fermions as a function of the dipole ratio $\muZ/\mug$ for a WIMP mass of $\mX = 130\GeV$ in the case of small splitting $\mX \approx \mXp$.

Surprisingly, when the dipole strength is normalized to yield an annihilation rate into $\gamma\gamma$ ($\gamma\ZZ$) in the range recently reported in refs.~\cite{Weniger:2012tx,Su:2012ft}, $0.3-1.3\times 10^{-27}{\rm cm^3/s}$, the annihilation rate at freeze-out into $f\bar{f}$ pairs near the expected ratio of $\muZ/\mug = -\tan\theta_{_W}$ is only a factor of $3-7$ ($5-10$) larger than the needed value of $6\times 10^{-26}{\rm cm^3/s}$. As we shall see in the next subsection this also yields a rate in direct detection experiments which can be probed with existing experiments and is surprisingly close to that reported by the DAMA collaboration for $\mX - \mXp \approx 100\keV$. Evidently, this surprising concordance is numerically not perfect, but is sufficiently interesting given the large astrophysical uncertainties.
  
\subsection{Direct Detection of \text{MiDM}: Constraining Cosmic Ray Gammas Underground}

A WIMP with a magnetic dipole can scatter against a nucleus' own magnetic dipole as well as its charge. In the case of elastic scattering the current limits from direct detection experiments on a WIMP of mass $\mX \approx 130\GeV$ and local mass density of $0.3\GeV/{\rm cm^3}$ are at the level of $\mu_X \lesssim 6 \times 10^{-5}\muN$ (see e.g.~\cite{Fortin:2011hv}). This excludes annihilation rates into di-photons at the level of $\sigma(\chi\chi\rightarrow \gamma\gamma) v \lesssim 10^{-34}~{\rm cm^3/s}$, far below anything we can hope to measure anytime soon. Thus, current direct detection limits robustly exclude the possibility of observing gamma ray lines from magnetic dipoles annihilations independently of the dipole strength. 

A more promising possibility is offered by MiDM~\cite{Chang:2010en} where the WIMP couples via a magnetic dipole to an excited state $\chi^{*}$. If the mass splitting is of order the kinetic energy of the WIMP in the halo $\Delta M = \mXp - \mX  \approx 100\keV$ then it may undergo inelastic scattering against the nucleus, but the corresponding rates in direct detection experiments are much reduced compared to the elastic scattering discussed in the last paragraph. Thus, the dipole strength can be larger and the annihilation rates of WIMPs into gamma rays considerably enhanced. This scenario then offers an interesting connection between observations in gamma ray lines and direct detection efforts that can be probed with current experiments. In particular, as discussed in ref.~\cite{Chang:2010en} it may explain the signal claimed by the DAMA collaboration\footnote{Recently, the KIMS collaboration has claimed to exclude the possibility by $O(1)$ at 90\% confidence \cite{Kim:2012rz}. However, this was a particular range of energies arising for a specific choice of quenching factors, both on NaI and CsI. When combined with the absence of a thorough discussion of energy resolutions (given that signal may leak into surrounding bins), the MiDM scenario appears intact, although likely requires $O(1)$ modulation fraction.}~\cite{Bernabei:2010mq} for WIMP masses of $\mX \sim 100\GeV$ and magnetic dipole strength in the range $\mu_{\chi} = 10^{-2} - 10^{-3} \muN$. Interestingly, for $\mX = 130\GeV$ and $\mu_\chi = 3\times 10^{-3}\muN$ the annihilation rate of WIMPs into gamma rays is $\sigma v(\chi\chi\rightarrow \gamma\gamma) \approx 10^{-27}{\rm cm^3/s}$ which can accommodate the excess recently reported in ref.~\cite{Weniger:2012tx,Su:2012ft}.

\subsection{Variations on a theme: Model Dependences in Indirect Signals}
As we laid out in eq. \ref{eq:scalingindirect}, the $\gamma$-ray line signals should be independent of the size of the dipole. The question arises as to how robust the size of the gamma ray signal will be to changes in the underlying model. Indeed, we have already seen examples of this: an MiDM model that {\em only} coupled to $\gamma$ would be a thermal relic with $\mu_\chi = \muther \sim 2 \times 10^{-3} \muN$, which leads to a $\gamma \gamma$ signal for a signal size of $\vev{\sigma v} \simeq 1.6 \times 10^{-28}{\rm cm^3 s^{-1}}$ (normalized to the case with $\Omega_\chi = \Omega_{DM}$). In contrast, in the presence of a dipole with the hypercharge gauge boson, the needed dipole for a thermal relic is roughly $\sqrt{3}$ smaller, leading to an effective signal size of $\vev{\sigma v} \simeq 2.5 \times 10^{-29}{\rm cm^3 s^{-1}}$. While the pure $\gamma$ dipole is close enough to explain the signal at 130 GeV, a dipole of hypercharge seems too small except in very cuspy halos. Thus, we should inquire whether there are any effects that modify this. As it turns out, there are at least two simple elements without enlarging the effective theory that can affect this.

\begin{figure*}[t,h]
\begin{center}
(a) \includegraphics[width=0.45 \textwidth]{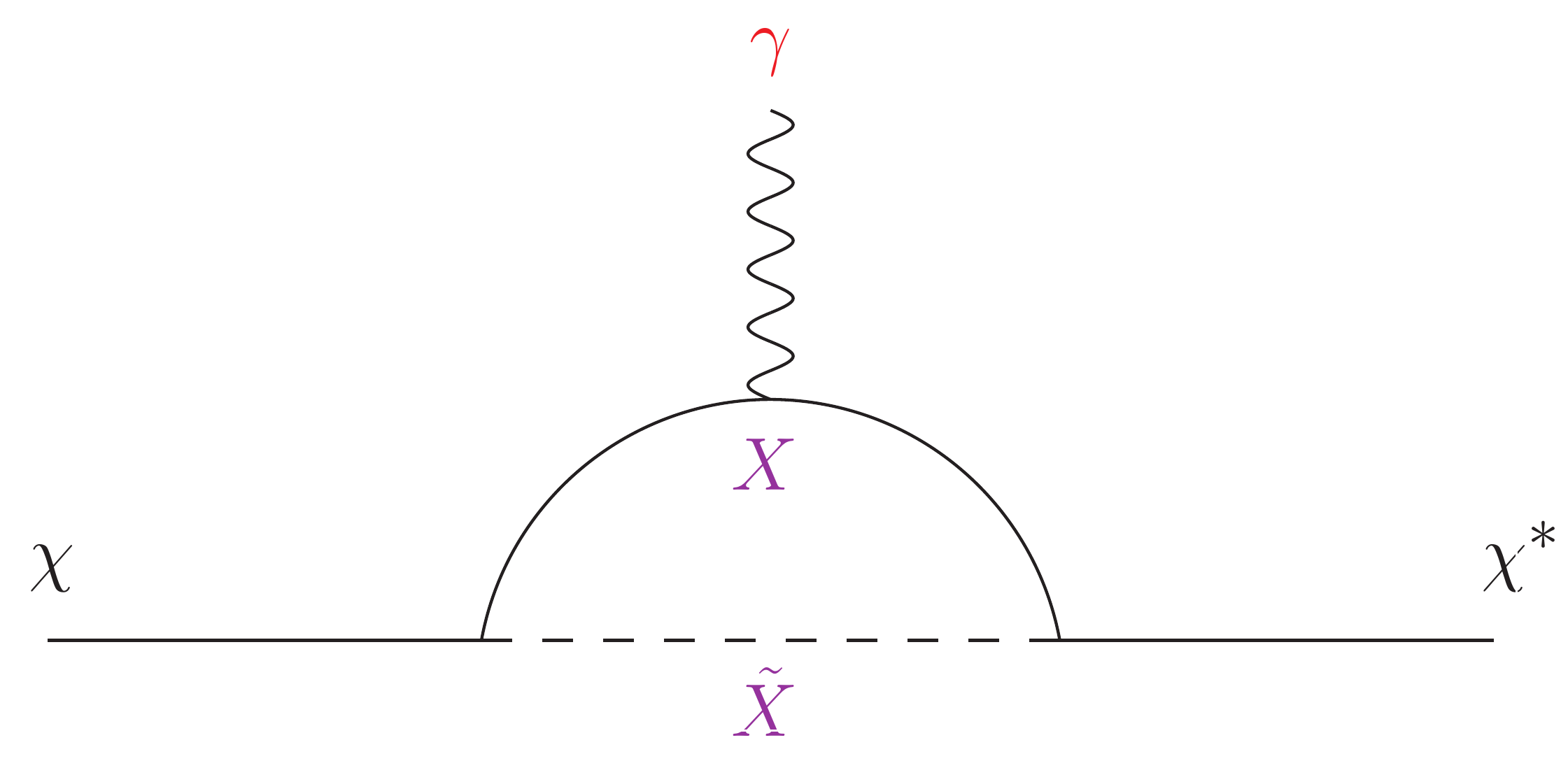}
(b)  \includegraphics[width=0.45 \textwidth]{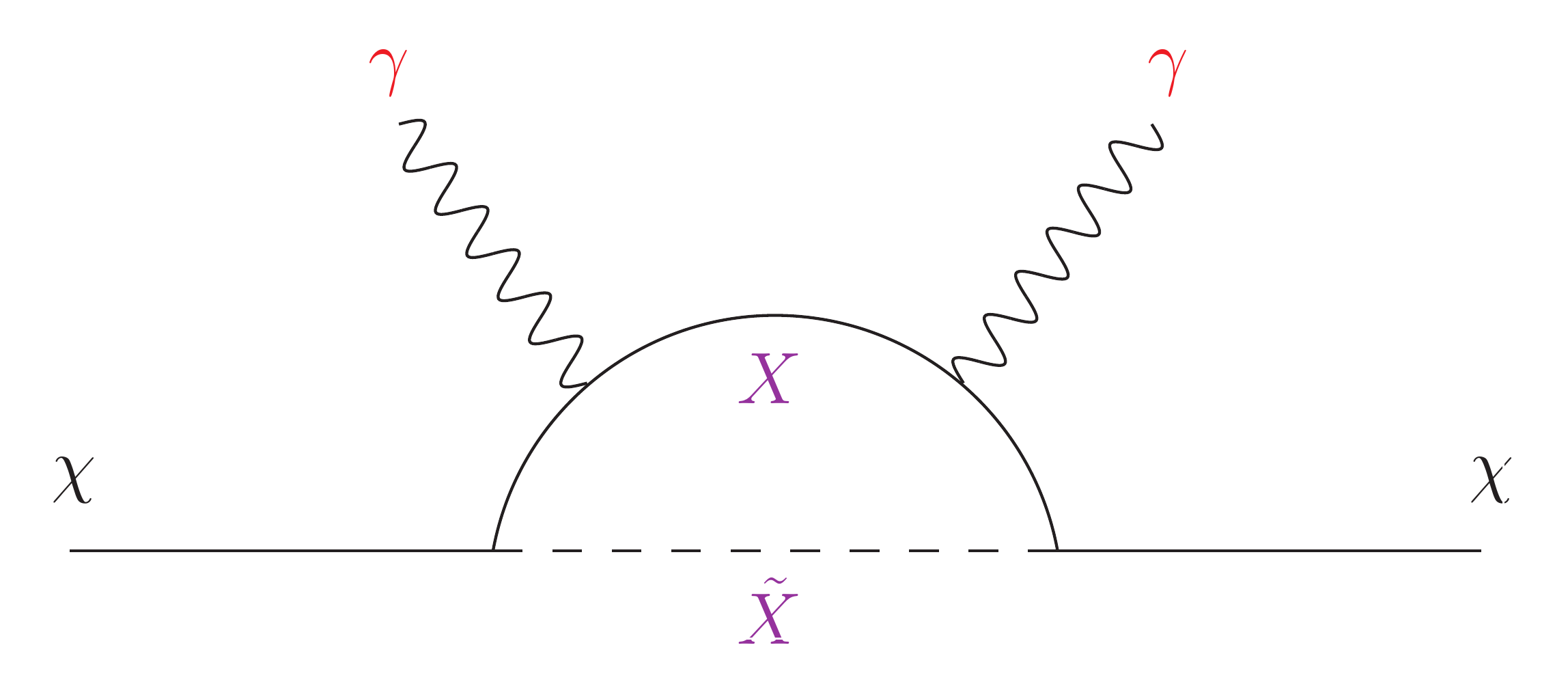}
\end{center}
\caption{({\em left}) Loop diagram contributing to dipole operator for MiDM model. ({\em right}) Comparable diagram contributing the Rayleigh operator.}
\label{fig:dipoleandrayleigh}
\end{figure*}

The first effect is the presence of form factors. Since the appropriate dipole scale is $\mu_\chi \gtrsim e/\tev$, it should be resolved near the WIMP mass scale. The annihilation into $\gamma \gamma$ samples a form factor with spacelike $q^2 =- m_\chi^2$, while the annihilation into charged pairs samples a form factor with timelike $q^2 = 4 m_\chi^2$. (The $\gamma Z$ signal samples form factors with $q^2 = \pm (m_\chi^2 - m_z^2/2)$.) If the form factor is being resolved at these momentum transfers, then treating it as a contact interaction is clearly wrong. Since $q^2$ for the s-channel diagram is four times larger than the t-channel, it is  reasonable that it could pick up a larger suppression factor. In this case the dipole would have to be increased by $\mu_\chi^2 = \muther^2/F^2(4 m_\chi^2)$. The resulting signal in $\gamma \gamma$ would be increased by $(F^2(-m_\chi^2)/F^2(4 m_\chi^2))^2$. We note that in this case, if the scale $\Lambda$ of new physics is not much higher than $m_\chi$, the Rayleigh operator we discuss in section \ref{sec:RayDM} may be generated as well, which could constructively interfere, enhancing the $\gamma \gamma$ rate.

It is worth dwelling on this last point for a moment. Generally, we should in fact expect the presence of the Rayleigh operator whenever the dipole operator is present. If the dipole is generated by a loop process as shown in Fig.~\ref{fig:dipoleandrayleigh}a then assuming a coupling $\lambda \chi X \tilde X$, the natural size for the dipole operator is
\be
g_Y \frac{\lambda^2}{16 \pi^2} \frac{1}{M_X}~.
\ee
By attaching a second external photon, the Rayleigh operator is also generated through the diagram shown in Fig.~\ref{fig:dipoleandrayleigh}b with a natural size
\be
g_Y^2 \frac{\lambda^2}{16 \pi^2} \frac{1}{M_X^3}~.
\ee
Thus, annihilation to $\gamma \gamma$ through the RayDM process is a one-loop process, while annihilation through the MiDM$^2$ process is effectively two-loop. Thus RayDM annihilation is relatively enhanced over MiDM$^2$ by a factor of
\be
\left(\frac{\lambda^2}{16 \pi^2} \right)^{-2}\frac{m_\chi^2}{M_X^2} ~.
\ee
Even for $m_\chi \sim M_X$, if the theory is at all perturbative, the Rayleigh contribution should dominate. If the Rayleigh contribution to the amplitude is even a few times larger than the MiDM$^2$ contribution, the size of the signal should be easily large enough to explain the 130 GeV signal. Thus, not only is it reasonable to believe that the Rayleigh operator could contribute, it should be a likely expectation.

A second possibility is the presence of CP violation. While both CP-conserving magnetic and CP-violating electric dipole moments produce s-wave $\gamma \gamma$ signals, only the s-channel diagram via a magnetic dipole yields an s-wave annihilation to charged fermions \cite{Fortin:2011hv}. Thus, with EDMs one can increase the present day $\gamma \gamma$ signal while producing only a $p$-wave suppressed annihilation at freezeout into $f \bar f$. With only EDMs, assuming a freezeout at $T\approx \mX/20$, the annihilation into gauge bosons dominates, and one has a signal cross section of $\approx 6 \times 10^{-26}~{\rm cm}^{3}{\rm s}^{-1}$ (as it is a Dirac fermion at freezeout), exceeding the Fermi limits. Thus, while pure EDM is excluded, a combination of EDM and MDM could produce the signal consistent with constraints. However, as we have stated, in the presence of such a large EDM, the direct detection limits would have excluded it unless the excited state is completely inaccessible. So while the first  possibility still offer the prospect of discovery at upcoming direct detection experiments, this second case seems unlikely to be found underground.

\section{$\text{RayDM}$}
\label{sec:RayDM}

\subsection{Searching for $\text{RayDM}$ in Gamma Rays}

The non-relativistic annihilation cross-section of RayDM into the different electroweak vector-bosons is sensitive only to the axial $\bar{\chi}\gamma_5\chi$ components to leading order in the velocity expansion. The differential cross-section is given by 
\be
\sigma(\chi\chi\rightarrow VV)v &=& \frac{ g_{_{VV}}^2 }{4\pi}  \frac{\mX^4 }{\LamR^6}~\mathcal{K}_{_{VV}},
\ee
with the kinematic functions $\mathcal{K}_{_{VV}}$ and couplings $g_{_{VV}}$ defined as
\vspace{5mm}
\begin{eqnarray}
\label{eqn:gVV}
&\mathcal{K}_{\gamma\gamma}& = 1, \quad \quad \quad \quad\quad\quad \quad  g_{\gamma\gamma} = \cX \cW^2 + \sX \sW^2, \\
&\mathcal{K}_{\rm \gamma {\scriptstyle Z}}& = 2\left(1-\frac{\mZ^2}{4\mX^2} \right)^{3},  \quad    g_{\rm \gamma {\scriptstyle Z}} = (\sX- \cX) \sW \cW, \\ 
&\mathcal{K}_{_{\rm ZZ}}& =  \left(1-\frac{\mZ^2}{\mX^2}\right)^{3/2}, \quad  g_{_{\rm ZZ}} = \cX \sW^2 + \sX \cW^2, \\
&\mathcal{K}_{_{WW}}& =  2 \left(1-\frac{\mW^2}{\mX^2}\right)^{3/2}, \quad  g_{_{WW}} =  \sX.
\end{eqnarray}
Here $\cX = \cos\thetaX$ and $\sX = \sin\thetaX$ and $\cW$ and $\sW$ are similarly defined with respect to the Weinberg angles. When $\mX$ is not too much smaller than $\LamR$ we expect some form-factor suppression to soften the behavior of this cross-section.

\begin{figure}
\begin{center}
\includegraphics[width=0.45 \textwidth]{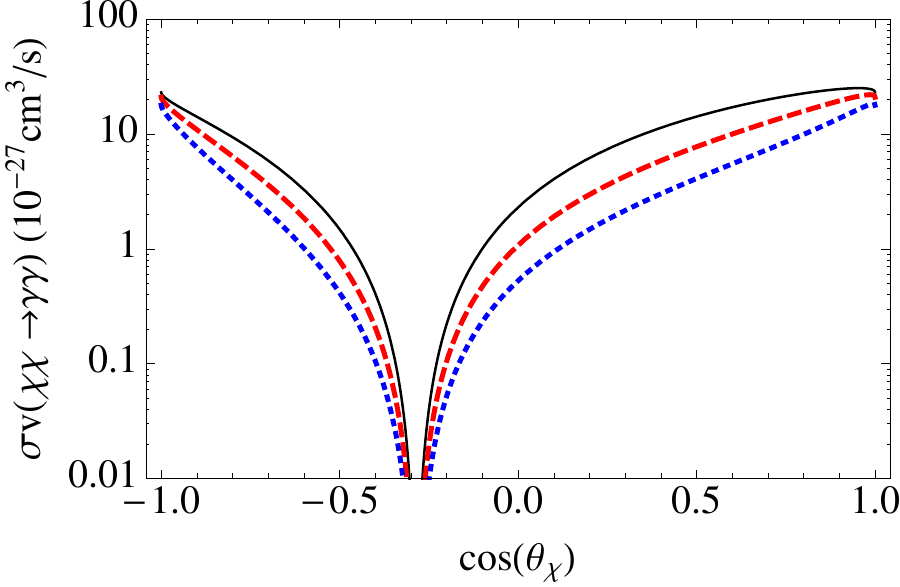}
\end{center}
\caption{The annihilation rate of WIMPs to di-photons, $\sigma(\chi\chi\rightarrow \gamma\gamma)$, as a function of the angle $\thetaX$ for different choices of the WIMP mass. For each mass choice the Rayleigh scale $\LamR$ is chosen so that the total annihilation cross-section yields the correct relic abundance. Shown are  $\mX = 100\GeV$ (solid-black), $\mX = 130\GeV$ (dashed-red). The dotted-blue curve depicts the asymptotic formula Eq.~(\ref{eqn:sigmagg}) which is independent of mass.}
\label{fig:ggXSvsTheta}
\end{figure}

In Fig.~\ref{fig:ggXSvsTheta} we plot the annihilation cross-section of $\chi\chi\rightarrow \gamma\gamma$ as a function of the WIMP mass for several values of $\cos\thetaX$, the relative coupling to the field-strengths in Eq.~(\ref{eqn:RDMLagrangian}). Requiring the right relic abundance, which for a Majorana fermion is obtained when the total annihilation cross-section at freeze-out is $3\times 10^{-26}{\rm cm^3/s}$, we can normalize the Rayleigh scale $\LamR$. For $\mX = 130\GeV$ this results in $\LamR = 440\GeV$ ($\LamR = 490\GeV$) in the case of $\cos\thetaX = 1$ ($\cos\thetaX = 0$). For a Dirac fermion the necessary annihilation cross-section is  $6\times 10^{-26}{\rm cm^3/s}$ and the Rayleigh scale is correspondingly a factor of $2^{1/6}$ smaller. The resulting gamma rays are monochromatic with $E_\gamma = \mX$. To qualitatively understand these results, we consider the limit where the WIMP mass is much heavier than the vector-bosons's. Then the expression for the total cross-section is particularly simple and by equating it to the required cross-section from relic abundance we can solve for $\LamR$ in terms of the WIMP mass and the angle $\thetaX$,
\be
\label{eqn:LamR}
\sum_{VV'}\sigma(&\chi\chi &\rightarrow VV')v = 3\times 10^{-26} {\rm cm^3/s} \\ \nonumber  &\Rightarrow& \LamR = 600\GeV\left(\frac{\mX}{200\GeV} \right)^{2/3}(2-\cos2\thetaX)^{\frac{1}{6}}.
\ee
With this value of the Rayleigh scale the annihilation rates into $\gamma\gamma$ and $\gamma\ZZ$ are
\be
\label{eqn:sigmagg}
\frac{\sigma(\chi\chi\rightarrow\gamma\gamma) v}{3\times 10^{-26}{\rm cm^3 s^{-1}}} &=& \frac{\left( \cW^2 \cX + \sW^2 \sX\right)^2}{2-\cos2\thetaX}\left(\frac{\LamR^{\rm th}}{\LamR} \right)^6,\\
\frac{\tfrac{1}{2}\sigma(\chi\chi\rightarrow\gamma\ZZ) v}{3\times 10^{-26}{\rm cm^3 s^{-1}}} &=& \frac{\cW^2\sW^2( \cX -  \sX)^2}{2-\cos2\thetaX}\left(\frac{\LamR^{\rm th}}{\LamR} \right)^6.
\ee
Here $\LamR^{\rm th}$ is the value of the Rayleigh scale that leads to the correct relic abundance. 
As can be expected when the Rayleigh operator is mostly associated with hypercharge the total cross-section is very close to the cross-section for annihilation into a photon pair $\sigma_{\rm tot} \approx \sigma(\chi\chi\rightarrow \gamma\gamma)$, which would result in too large a signal\footnote{Although, as we discuss in the conclusions, a subdominant DM component would plausibly give the right signal.}. When the Rayleigh operator is mostly associated with the non-abelian $\SUWeak$ part, the annihilation into photons is suppressed compared with the total cross-section due to the Weinberg angle and the $\chi\chi\rightarrow W^+W^-$ channel. More quantitatively, in the pure $\SUWeak$ case, $\sigma_{gg}/\sigma_{tot} \approx 1/28$ and $1/2 \times \sigma_{\gamma Z}/\sigma_{tot} \approx 1/12$. The photon-to-hadron ratio $(\sigma_{\gamma \gamma}+1/2 \sigma_{\gamma Z})/(\sigma_{tot}-\sigma_{\gamma \gamma}-1/2 \sigma_{\gamma Z})\approx 1/7.6$. In the case of hypercharge RayDM the equivalent numbers are $\sigma_{gg}/\sigma_{tot} \approx 1/1.4$,  $1/2 \times \sigma_{\gamma Z}/\sigma_{tot} \approx 1/7$. The photon-to-hadron ratio is $\approx 5$, so no significant hadronic emission is present in this case.

\begin{figure}
\begin{center}
\includegraphics[width=0.46 \textwidth]{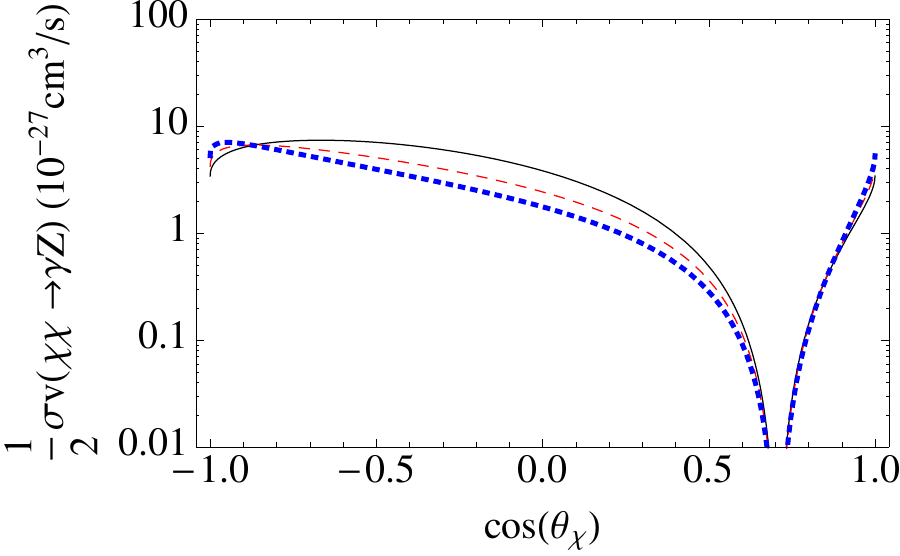}
\includegraphics[width=0.45 \textwidth]{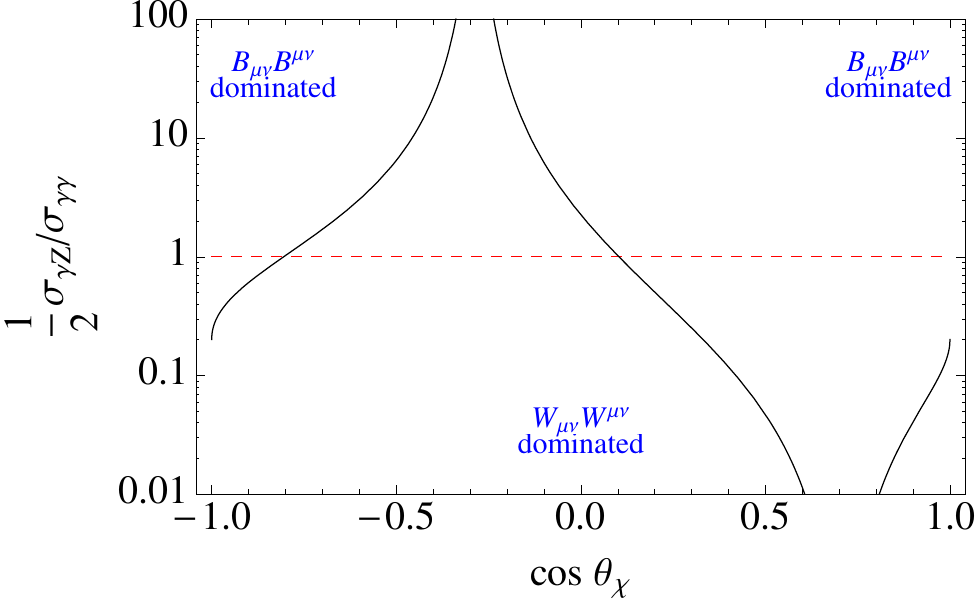}
\end{center}
\caption{The top pane is similar to Fig.~\ref{fig:ggXSvsTheta} above, but for half the annihilation rate into a photon and a $\ZZ$ boson, $\tfrac{1}{2} \sigma(\chi\chi\rightarrow \gamma Z)$. In the bottom pane we plot the ratio of the annihilation rate into $\gamma\ZZ$ to that into $\gamma\gamma$ for $\mX = 130\GeV$ as a function of the angle $\thetaX$.}
\label{fig:gZXSvsTheta}
\end{figure}

An additional process contributing to monochromatic gamma ray signal is of course $\chi\chi\rightarrow \gamma {\rm Z}$ with a lower energy of $E_\gamma = \mX-\mZ^2/4\mX$.  In Fig.~\ref{fig:gZXSvsTheta} we plot the annihilation rate associated with this channel as well as its ratio to the di-photon rate. Depending on the DM halo profile and the angle $\cos\thetaX$, both the $\gamma\gamma$ and $\gamma\ZZ$ rates are in the interesting range reported in ref.~\cite{Weniger:2012tx},  $3\times 10^{-28} - 2\times 10^{-27}{\rm cm^3/s}$. This points to a fairly low Rayleigh scale of $\LamR \approx 500\GeV$.

One might worry about the validity of this picture given that the Rayleigh scale is rather low. We come back to this point in section~\ref{sec:collider} where this issue is particularly important, however, for the purpose of non-relativistic annihilation it is only necessary for the Rayleigh scale to be somewhat larger than the WIMP mass. Nevertheless, since the WIMP mass is not much lower than the Rayleigh scale, it may be appropriate to include a form-factor. Thinking about RayDM as MiDM$^2$ allows to resolve the 4-point interaction with the exchange of the excited state $\chi^*$. Consulting the corresponding annihilation rates in MiDM, eqs.~(\ref{eqn:MiDManngg}-\ref{eqn:MiDMannZZ}) we see for example that the process $\chi\chi\rightarrow \gamma\gamma$ is diminished by a factor of $(1+\mX^2/\mXp^2)^{-2}$.

\subsection{Direct Detection of RayDM}

\begin{figure}
\begin{center}
\includegraphics[width=0.4 \textwidth]{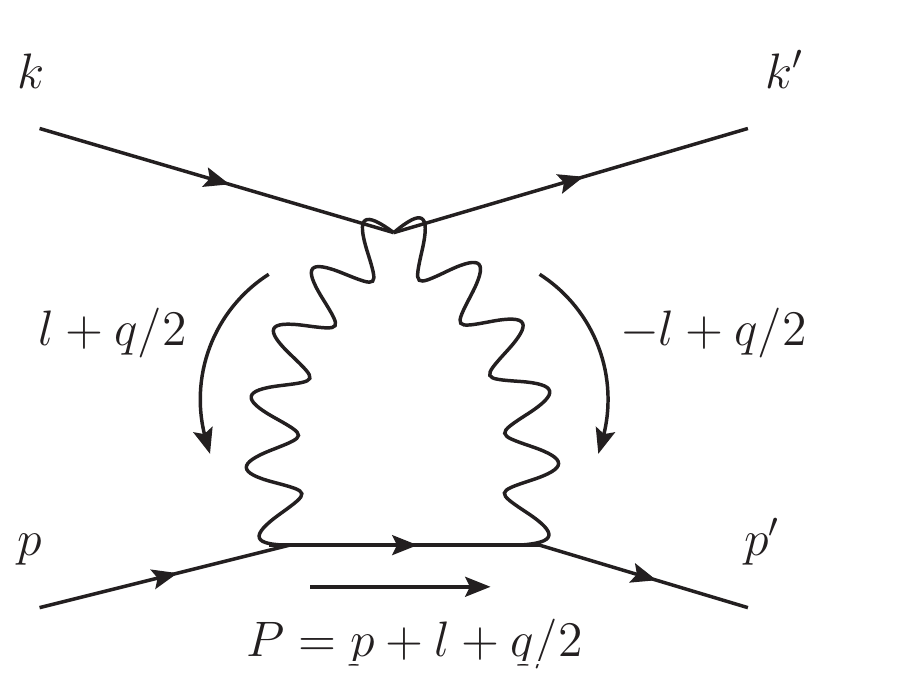}
\end{center}
\caption{Elastic scattering of RayDM against the nucleus through two photons exchange.}
\label{fig:elastic_scattering}
\end{figure}

The scattering of RayDM against matter is complicated by the fact that at least two force mediators have to be exchanged. The exchange of two photons leads to the least amount of suppression and so we concentrate on this case and consider the Lagrangian
\be
\mathcal{L} = \frac{g_{\gamma\gamma}}{4\LamR^3}~\bar{\chi}\chi~ F_{\mu\nu}F^{\mu\nu}.
\ee 
Two distinct processes are possible: $\chi N \rightarrow \chi N$ elastic scattering through the loop shown in Fig.~\ref{fig:elastic_scattering}; $\chi N \rightarrow \chi N +\gamma$ tree-level scattering. The latter channel is extremely suppressed due to phase-space. The total cross-section for the elastic channel was previously calculated in the thorough work of ref.~\cite{Pospelov:2000bq} in the approximation that the nucleus is much heavier than the WIMP by considering the scattering of the WIMP off the external electric field generated by the nucleus. In appendix~\ref{app:eval_loop_int} we provide a different derivation which leads to a  more exact result that is valid when the WIMP mass cannot be neglected relative to that of the nucleus. At leading order in the velocity expansion the amplitude for this process is given by
\be
\label{eqn:ESAmp}
\hspace{-7mm}i\mathcal{M} =  \frac{i \alpha Z^2g_{\gamma\gamma}}{4}\frac{Q_0}{\LamR^3}~\mathcal{F}\left( \frac{\left|{\bf q}\right|^2}{Q_0^2}\right) \bar{u}(k') u(k)~ \bar{u}(p') u(p),
\ee
where $Z$ is the nucleus charge, $\alpha = 137^{-1}$ is the fine-structure constant, $Q_0$ is  the nuclear coherence scale, and the momentum transfer is related to the relative velocity between the WIMP and the nucleus and the angle of scattering $\theta$ in the centre of mass frame through $\left| {\bf q}\right|^2 = 2\mu^2v^2(1-\cos\theta)$. The function $\mathcal{F}(x)$ decreases exponentially for high momentum transfers and is of order unity near the origin, $\mathcal{F}(0) = 2/\sqrt{\pi}$. The spin-independent differential cross-section in the centre of mass frame is then,
\be
\label{eqn:ESdiffXS}
\frac{d\sigma}{d\cos\theta} = \frac{\mu_{\chi N}^2}{2\pi}\left|\frac{\alpha Z^2g_{\gamma\gamma}}{4}\frac{Q_0}{\LamR^3}~\mathcal{F}\left( \frac{\left|{\bf q}\right|^2}{Q_0^2}\right) \right|^2.
\ee
Here $\mu_{\chi N}$ is the nucleus-WIMP reduced mass. To a good approximation we can use $\mathcal{F}\left(\left|{\bf q}^2\right|/Q_0^2\right) \approx \mathcal{F}(0)$ and so the total spin-independent cross-section \textit{per nucleon} is given by
\be
\sigma_p^{\rm SI} \approx \frac{\alpha^2 Z^4 g_{\gamma\gamma}^2}{4\pi^2 {\rm A}^4}~ \frac{\mN^2 Q_0^2}{\LamR^6},
\ee
where ${\rm A}$ is the nucleon number. This is an extremely small cross-section for an electroweak scale WIMP. For example, taking the nuclear coherence scale $Q_0 = \sqrt{6}\left(0.3 + 0.89 {\rm A}^{1/3} \right)^{-1}$, the Rayleigh scale $\LamR = 500\GeV$, and setting $g_{\gamma\gamma}= 1$ yields $\sigma_p^{\rm SI} \approx 10^{-49}~{\rm cm}^2$ for scattering on xenon. If the Rayleigh scale is considerably lower, $\LamR \lesssim 100\GeV$ then the scattering rates become appreciable. That requires much lighter WIMPs than what we set to explore in this work and we leave it for a future study to elucidate the prospects associated with this part of parameter space. 


\section{Collider Phenomenology}
\label{sec:collider}

In MiDM, the production mode in colliders is simply $f\bar{f}\rightarrow \chi\chi^*$ followed by the prompt decay of the heavier state to a photon or a $\ZZ$ boson, as shown in Fig.~\ref{fig:XXprodMiDM}. As we emphasized throughout this paper, the phenomenology of MiDM depends crucially on the mass difference between the WIMP $\chi$ and the heavier state $\chi^*$ and its collider phenomenology is no exception. When the mass splitting is small searching for MiDM in collider proceeds in the same fashion as searching for other WIMPs, namely by looking for mono-jet or mono-photon events from an unbalanced initial state radiation (see for example the excellent recent works of~\cite{Bai:2010hh,Goodman:2010yf,Goodman:2010ku}). When the mass splitting is large the collider signatures of MiDM are more similar to that of RayDM. 

In RayDM, since the coupling to WIMPs  requires at least two vector-bosons, the collider signatures are somewhat different than usual. There are two different processes that may be searched for: $f\bar{f} \rightarrow f\bar{f} \chi\chi$ through a vector-boson or photon fusion; and $f\bar{f} \rightarrow \chi\chi V$, through an intermediate vector-boson, where $V = \gamma,{\rm Z}$, or $W^\pm$. This last process, shown in Fig.~\ref{fig:XXprodRDM},  enjoys a larger cross-section and results in the production of a photon or an electroweak vector-boson in association with large missing energy.  We therefore concentrate on this possibility below. We begin by discussing the MiDM scenario, followed by RayDM, and finally discuss the actual constraints. General formulas for the differential cross-sections in the different cases are given in appendix~\ref{app:collider_production}. 

\subsection{MiDM}

In MiDM the dominant mode is the production of the heavier state in association with the WIMP through $f\bar{f}\rightarrow \gamma/\ZZ\rightarrow \chi\chi^*$. After production, the heavier state subsequently decays to the WIMP through the emission of a photon or a vector-boson as shown in Fig.~\ref{fig:XXprodMiDM}. The differential cross-section for this process is given in Eq.~(\ref{eqn:MiDMprodXS}) in the appendix. 

When the splitting between the excited state and the ground state is much smaller than the mass $\DeltaM \ll \mX$ the resulting photon or vector-boson is too soft to be searched for directly. Ref.~\cite{Bai:2011jg} proposed some interesting ideas for looking for iDM in colliders when the mass splitting is in the GeV range and the heavier state decays into pions. More generally and without reliance on such specialized techniques,  the collider phenomenology in this case is identical to the usual case of WIMP pair production, but through a dipole operator. It can be searched for in a general way by tagging on initial state radiation. This  was nicely worked out in ref.~\cite{Fortin:2011hv} for mono-jet searches in the case of degenerate states ($\mXp=\mX$) for both magnetic as well as electric dipoles. When the splitting is of order the mass and larger, the emitted photon or vector-boson may be sufficiently energetic to be searched for directly. This can be searched for without reliance on initial state radiation hence enjoying a larger cross-section. In this case collider searches for mono-photons place strong constraints on this scenario as discussed below in the final part of this section. 

\begin{figure}
\begin{center}
\includegraphics[width=0.45 \textwidth]{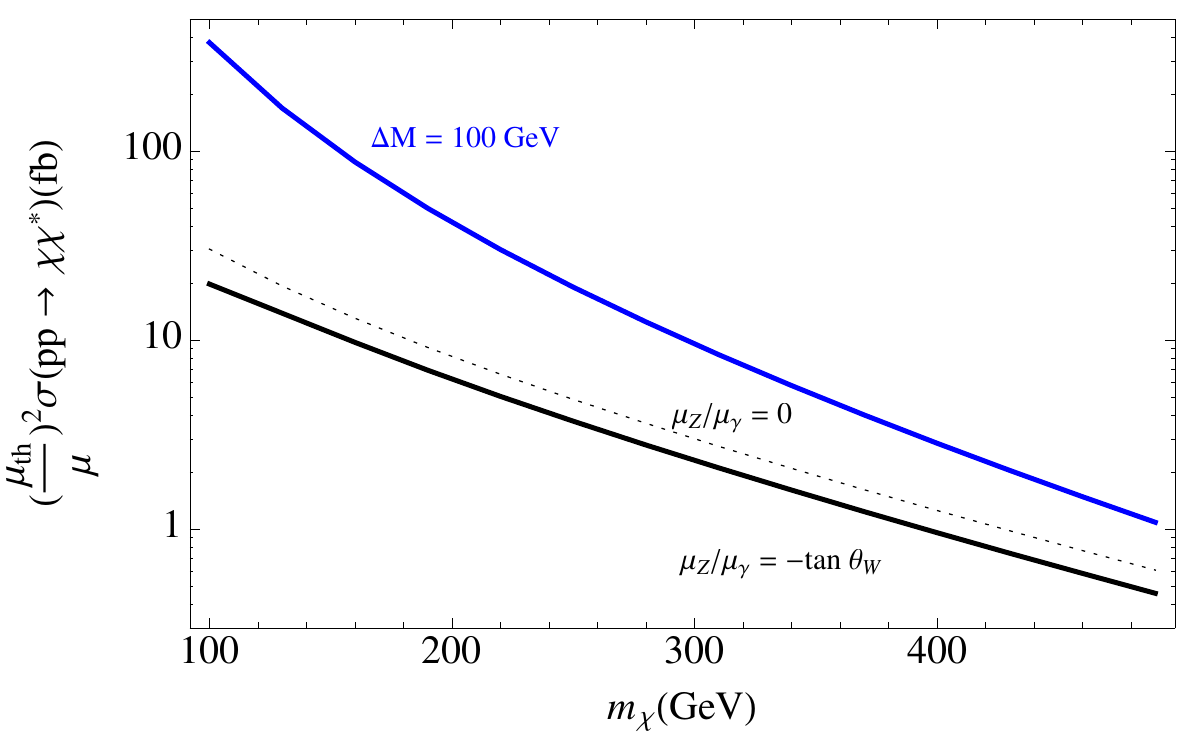}
\end{center}
\caption{Minimal production cross-section for MiDM, $pp\rightarrow \chi\chi^*$, through $\gamma/\ZZ$ in the s-channel against the mass of the WIMP. The lower (black) curves show the case of small splitting $\mX\approx\mXp$ for two choices of the dipoles' ratio $\muZ/\mug = -\tan\theta_{_W}$ (solid) and $\muZ/\mu = 0$ (dotted). The dipole strengths in this case are normalized to yield the correct relic abundance when the annihilation is dominantly into fermion pairs, Eq.~(\ref{eqn:MiDMannffbar}). While the $\gamma\gamma$ signal remains invariant as the dipole strength increases, the collider cross-section will increase as $\mug^2$. The upper blue curve depicts the case when the mass splitting is large $\mXp-\mX = 100\GeV$ for $\muZ/\mug = -\tan\theta_{_W}$. The dipole strength is normalized to yield the correct relic abundance when the annihilation is dominantly into vector-boson pairs, Eqs.~(\ref{eqn:MiDManngg}-\ref{eqn:MiDMannZZ}). }
\label{fig:MiDM_production}
\end{figure}

In Fig.~\ref{fig:MiDM_production} we plot the production cross-section for MiDM as a function of the WIMP mass for several choices of the parameters and the mass splittings. We recall that in the case of small splitting where the relic abundance is determined by the annihilation into fermions the $\gamma\gamma$ signal is independent of the dipole strength whereas the collider cross-section scales as the square of the dipole. Thus, the cross-section shown in Fig.~\ref{fig:MiDM_production} should be interpreted as the minimal cross-sections when the dipole strength is normalized to yield the correct relic abundance, $\mu = \muther$. For the same reason, the ratio $\frac{\sigma\left(pp\rightarrow \gamma/Z\rightarrow \chi\chi^*\right)^2}{\sigma v (\chi\chi \rightarrow \gamma\gamma)}$ is independent of the dipole strength for a given mass and choice of $\muZ/\mug$.  It is given in Table~\ref{tbl:productionVSannihilation} and allows for a straightforward comparison between the rates in astrophysical processes and the cross-sections in colliders.

{ \renewcommand{\arraystretch}{1.4}
\begin{table}[htdp]
\begin{center}
\begin{tabular}{|c|c|c|}
\hline
& $\muZ/\mug = -\tan\theta_{_W}$ & $\muZ/\mug = \cot\theta_{_W}$ \\
\hline
$(\mX,\Delta M)$ & & \\
$(130\GeV, 0)$ & \quad $\frac{\left( 28~ {\rm fb}\right)^2}{10^{-28}~{\rm cm^3 s^{-1}}}  \quad $ & \quad $\frac{\left( 0.4~ {\rm pb}\right)^2}{10^{-28}~{\rm cm^3 s^{-1}}}  \quad $ \\
$(130\GeV, 100\GeV)$ &  $\frac{\left( 16~ {\rm fb}\right)^2}{10^{-28}~{\rm cm^3 s^{-1}}}  $ & $\frac{\left( 0.2~ {\rm pb}\right)^2}{10^{-27}~{\rm cm^3 s^{-1}}}  $\\  
$(300\GeV, 0)$ &  $\frac{\left( 1.6~ {\rm fb}\right)^2}{10^{-28}~{\rm cm^3 s^{-1}}}  $ & $\frac{\left( 23~ {\rm fb}\right)^2}{10^{-28}~{\rm cm^3 s^{-1}}}  $\\  
$(300\GeV,100\GeV)$ &  $\frac{\left( 1.1~ {\rm fb}\right)^2}{10^{-28}~{\rm cm^3 s^{-1}}}  $ & $\frac{\left( 15~ {\rm fb}\right)^2}{10^{-28}~{\rm cm^3 s^{-1}}}  $\\  
\hline
\end{tabular}
\end{center}
\caption{For a given WIMP mass $\mX$ and splitting $\Delta M = \mXp-\mX$ the ratio $\frac{\sigma\left(pp\rightarrow \gamma/Z\rightarrow \chi\chi^*\right)^2}{\sigma v (\chi\chi \rightarrow \gamma\gamma)}$ is independent of the dipole strength and is displayed in this table for several choices of the masses and couplings. The production cross-section is calculated at leading order for the LHC with $\sqrt{s} = 7\TeV$. }
\label{tbl:productionVSannihilation}
\end{table}%
}

\subsection{RayDM}

Collider searches for dark matter are inevitably tied up in the embedding of dark matter into a complete theory and RayDM is no exception. 
In this subsection we discuss the possibility of directly producing WIMPs in colliders in the RayDM scenario. As we shall see, the phenomenology is sensitive to the UV physics that resolves the Rayleigh operator and as a result is more model dependent. This is in contrast to the phenomenology associated with direct and indirect detection efforts discussed in the previous sections which is insensitive to the details of the UV physics. It is important to keep this contrast in mind when considering the impact of collider constraints on the previous sections (this issue also arises in other models of DM, see~\cite{Shoemaker:2011vi} and references therein for some related discussion). 

In the case of RayDM, the differential cross-section for $f\bar{f} \rightarrow \bar{\chi}\chi \gamma$ at center of mass energy $\sqrt{s}\gg \mX$ through an intermediate $\gamma/\ZZ$  is given by
\be
\frac{1}{\sigma_{\rm tot}}\frac{d\sigma}{dp_T} = \frac{20~p_T}{s}\left(1-\frac{4~p_T^2}{s}\right)^{3/2},
\ee
where $p_T$ is the transverse momentum of the photon in the center-of-mass frame, and
\be
\label{eqn:ffbartotxs}
&\sigma_{\rm tot} &\left( f\bar{f}\rightarrow \bar{\chi}\chi\gamma \right) = \frac{\alpha~ q_f^2}{3840 \pi^2} ~\frac{s^2}{\LamR^6} \\ \nonumber &\times& \left(g_{\gamma\gamma}^2 + 2 g_{\gamma\gamma} g_{\gamma\ZZ} v_f \xi(s) + g_{\gamma\ZZ}^2 (v_f^2+a_f^2) \xi^2(s)\right) .
\ee
Here $q_f$ is the fermion's electric charge, $v_f$ ($a_f$) is the ratio of its vector (axial-vector) coupling the Z boson to its electromagnetic coupling, and $\xi(s) = s/(s-\mZ^2)$. We note that the transverse momentum distribution is such that most photons are fairly central, which is important for the mono-photon searches. Similar relations can be obtained for the production cross-section for $f\bar{f}\rightarrow \bar{\chi}\chi W $. In the limit where the WIMP mass and the W-boson mass are both much smaller than the incoming center of mass energy an identical distribution in $p_T$ results. This motivates mono-W searches, looking for the final state $\Wpm$ produced in association with the invisible $\chi\chi$ pair.  

At LEP for example, where $\sqrt{s} \approx 200\GeV$, one obtains $\sigma_{tot} \approx 7 \times 10^{-3}~{\rm fb} ~\left(500\GeV/\LamR \right)^6$ for $\thetaX = 0$. This cross-section is much too low unless the Rayleigh scale is brought down considerably. This is in good qualitative agreement with the very thorough investigation of ref.~\cite{Kathrein:2010ej} where bounds on unparticle production at LEP were presented\footnote{The Rayleigh operator has scaling dimension of $\Delta = 3$ in the notation of ref.~\cite{Kathrein:2010ej}, which was not considered by the authors for good reasons. For $\Delta = 2$ they find that the unparticle scale can be as low as $190\GeV$. Attempting to extrapolate to $\Delta = 3$ is not very useful since the energy available is greater than the cut-off scale and some UV completion is needed to resolve the non-renormalizable Rayleigh operator.}.

At Tevatron and LHC, one must convolve the above expressions against the parton luminosity functions. The resulting cross-section is larger, but for those values of the Rayleigh scale where the cross-section is sufficiently large to be interesting the theory requires a UV completion. One possible UV completion of RayDM is of course MiDM in the case when the mass splitting is very large. Integrating out the excited state $\chi^*$ one recovers the RayDM interactions. So schematically
\be
{\rm MiDM}^2 \overset{\mXp \gg E}{\xrightarrow{\hspace*{1.3cm}}} {\rm RayDM}.
\ee
In this case the Rayleigh scale $\LamR$ is connected with the magnetic dipole $\mu_{\chi}$ through $\LamR^3 = \mXp/2\mu_{\chi}^2$ and one can easily translate the results for MiDM in the previous subsection to the case of RayDM.

\begin{figure}
\begin{center}
\subfigure[~ MiDM]{
\label{fig:XXprodMiDM}
\includegraphics[width=0.3 \textwidth]{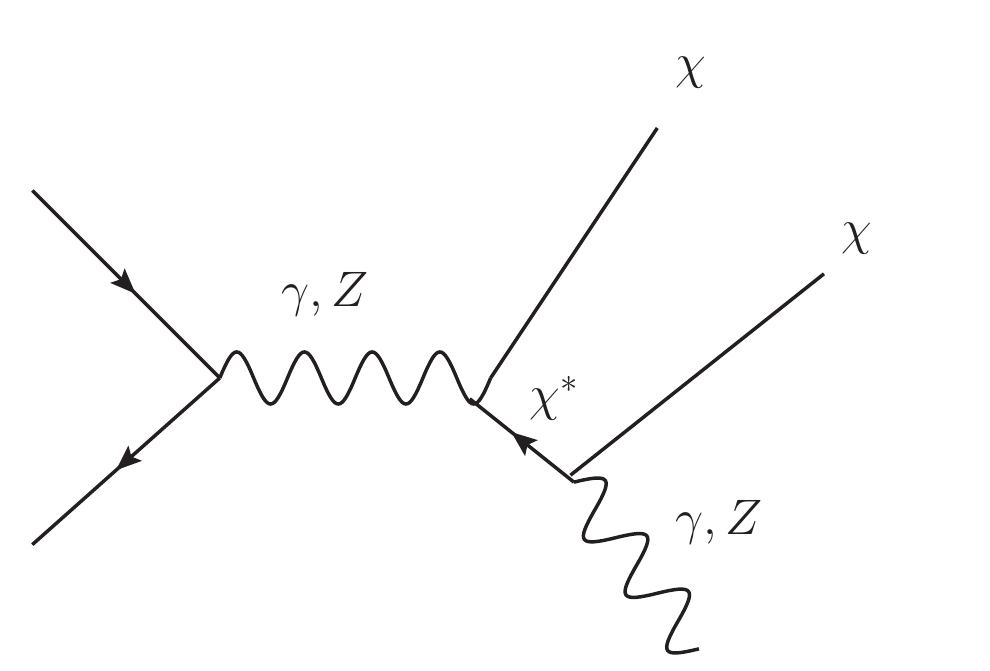}
} 
\subfigure[ ~RayDM]{
\label{fig:XXprodRDM}
\includegraphics[width=0.3 \textwidth]{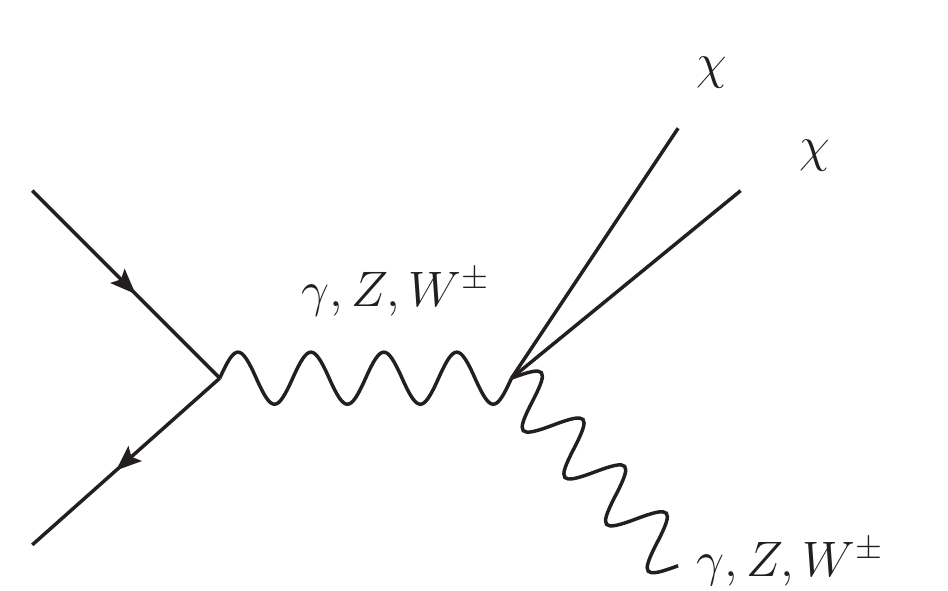}
} 
\subfigure[ ~$\phi FF$]{
\label{fig:XXprodphiFF}
\includegraphics[width=0.3 \textwidth]{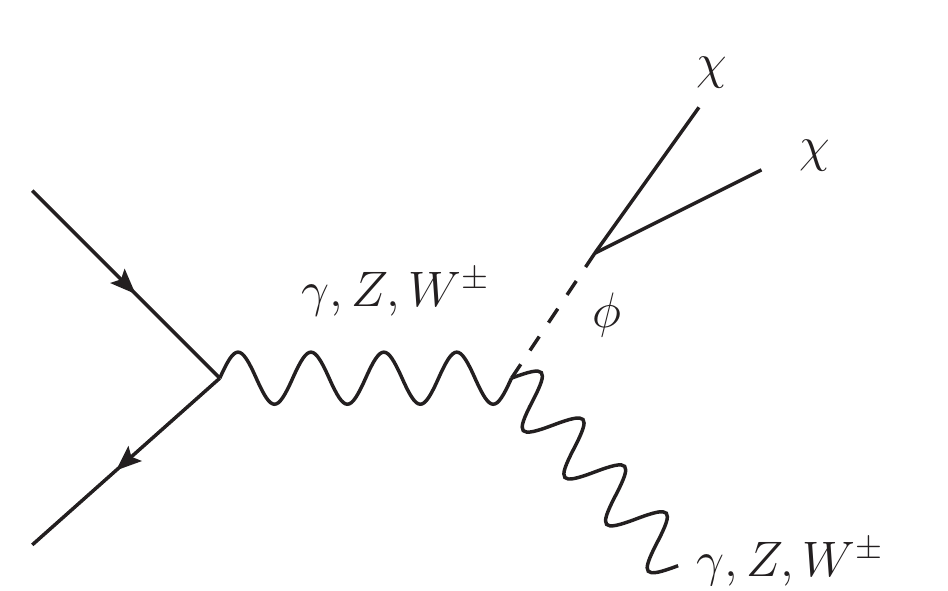}
} 
\end{center}
\caption{The production of WIMPs at colliders is shown in Fig.~\ref{fig:XXprodRDM} for the RayDM scenario. When the Rayleigh scale is comparable or lower than the energies involved in the collision, the Rayleigh operator must be resolved. In Fig.~\ref{fig:XXprodMiDM} we show the corresponding process in the case where RayDM is the result of integrating out a heavy excited state in MiDM. In Fig.~\ref{fig:XXprodphiFF} we show the the process in the case where the Rayleigh operator is resolved in terms of a new scalar. More details on the UV completions are provided in the text. }
\label{fig:XXproduction}
\end{figure}

Another possible UV completion involves a scalar $\phiS$ and a pseudoscalar $\phiA$, which couple directly to the WIMP as well as to the field-strengths of ${\rm U_Y(1)}$ and $\SUWeak$. We parametrize this theory with
\be
\label{eqn:HeavyScalarsLag}
\mathcal{L} &=& \tfrac{1}{2}\partial_\mu \phiS \partial^\mu\phiS - \tfrac{1}{2}m_{\phiS}^2 \phiS^2 + \tfrac{1}{2}\partial_\mu \phiA \partial^\mu\phiA - \tfrac{1}{2}m_{\phiA}^2 \phiA^2 \\ \nonumber &+& \phiS \bar{\chi}\chi + \phiA \bar{\chi}\gamma_5 \chi + \frac{\cos\thetaX}{4\Lambda_{_{\rm UV}}}\phiS B_{\mu\nu}B^{\mu\nu}+ \frac{\cos\thetaX}{4\Lambda_{_{\rm UV}}}\phiA B_{\mu\nu}\tilde{B}^{\mu\nu} \\ \nonumber &+&   \frac{\sin\thetaX}{4\Lambda_{_{\rm UV}}}\phiS {\rm Tr} W_{\mu\nu}W^{\mu\nu}+ \frac{\sin\thetaX}{4\Lambda_{_{\rm UV}}}\phiA {\rm Tr} W_{\mu\nu}\tilde{W}^{\mu\nu}.
\ee
Here we have taken the Yukawa couplings of the scalars to the WIMP to be order unity and used $\Lambda_{_{\rm UV}}$ to denote the scale of the dimension-5 operators. Integrating out the scalars, $\phiS$ and $\phiA$ we generate the different operators of RayDM. When the scalars are light enough to be produced in colliders the dominant process shown in Fig.~\ref{fig:XXprodphiFF} is $f\bar{f} \rightarrow V\phiS(\phiA)$ followed by the decay of $\phiS$ ($\phiA$) to a WIMP pair. We note that this example in fact results in a more general version of RayDM where the relative coupling of the scalar $\bar{\chi}\chi$ and pseudoscalar $\bar{\chi}\gamma_5\chi$ to the field strengths is arbitrary. This is important in the case of comparing direct detection rates (which are sensitive to the scalar piece) to indirect detection rates (which are sensitive to the pseudoscalar piece). This distinction does not play an important role for the purpose of collider phenomenology~\cite{Bai:2010hh,Goodman:2010yf,Goodman:2010ku}.

\begin{figure}
\begin{center}
\includegraphics[width=0.45 \textwidth]{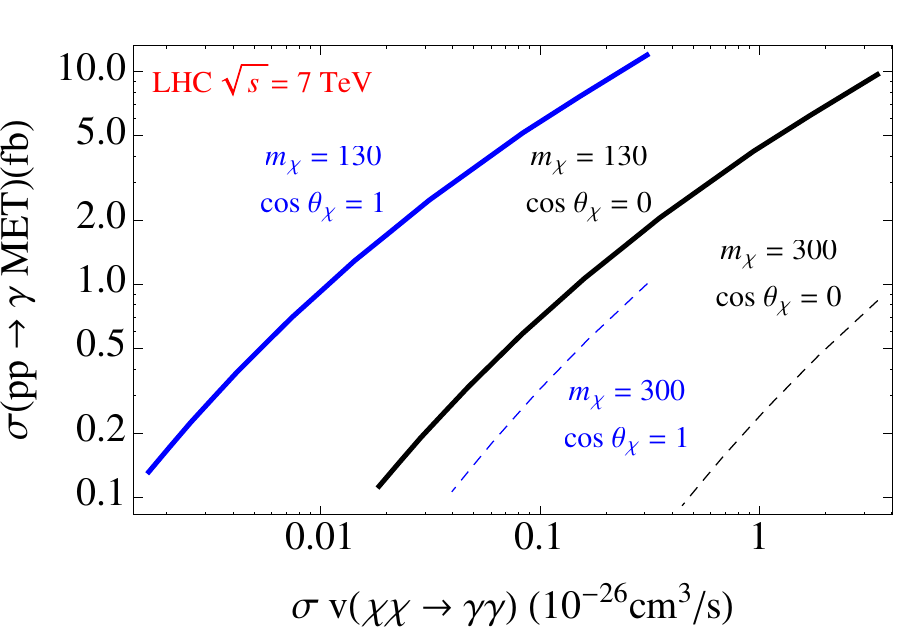}
\end{center}
\caption{A plot of the production cross-section for the heavy scalars in association with a photon, $pp\rightarrow \gamma \phiS (\phiA)$ against the annihilation rate of WIMPs to two photons. The mass of the scalar is allowed to vary between $2\mX$ and $\Lambda_{\rm UV}$. The cut-off scale is set to $\Lambda_{\rm UV} = \TeV$, but since both the production cross-section as well as the annihilation rate scale as $\Lambda_{\rm UV}^{-2}$ any other choice can be obtained with a simple rescaling. The solid (dashed) curves correspond to $\mX = 130\GeV$ ($\mX = 300\GeV$) whereas the blue (black) curves correspond to $\cos\thetaX =1$ ($\cos\thetaX =0$).  }
\label{fig:phi_production}
\end{figure}

In this case, the distribution of the transverse momentum of the photon in the center of mass frame is given by
\be
\label{eqn:ffTOphigammaDiff}
\frac{1}{\sigma_{\rm tot}}\frac{d\sigma}{dp_T} = \frac{3~p_T}{s} \frac{\left((1 - \frac{m_{\phiS}^2}{s})^2 + \frac{2p_T^2}{s}\right)}{\left(1- \frac{m_{\phiS}^2}{s} \right)^3 \sqrt{(1 -  \frac{m_{\phiS}^2}{s})^2 - \frac{4p_T^2}{s}}}.
\ee
We note that this function is strongly peaked towards the kinematical limit $p_T^{\rm (max)} = \frac{\sqrt{s}}{2} \left(1-m_{\phiS}^2/s\right)$. This is in sharp contrast to typical mono-photon signatures of dark matter production where the photon originates from initial state radiation and hence its $p_T$ is dominantly soft. Here the total cross-section is given by
\be
\label{eqn:ffTOphigammaTot}
&\sigma_{\rm tot} &\left( f\bar{f}\rightarrow \bar{\chi}\chi\gamma \right) = \frac{\alpha~ q_f^2}{24\Lambda_{\rm UV}^2} ~\left(1 - \frac{m_{\phiS}^2}{s}\right)^3 \\ \nonumber &\times& \left(g_{\gamma\gamma}^2 + 2 g_{\gamma\gamma} g_{\gamma\ZZ} v_f \xi(s) + g_{\gamma\ZZ}^2 (v_f^2+a_f)^2 \xi^2(s)\right) .
\ee
\vspace{1mm}
The couplings and the function $\xi(s)$ are defined after Eq.~(\ref{eqn:ffbartotxs}) above. Similar expressions hold for the axial scalar $\phiA$. In Fig.~\ref{fig:phi_production} we plot the production cross-section of a photon in association with one of the scalars against the annihilation rate of WIMP into two photons.  

\subsection{Limits from Colliders}

\begin{figure}
\begin{center}
\includegraphics[width=0.45 \textwidth]{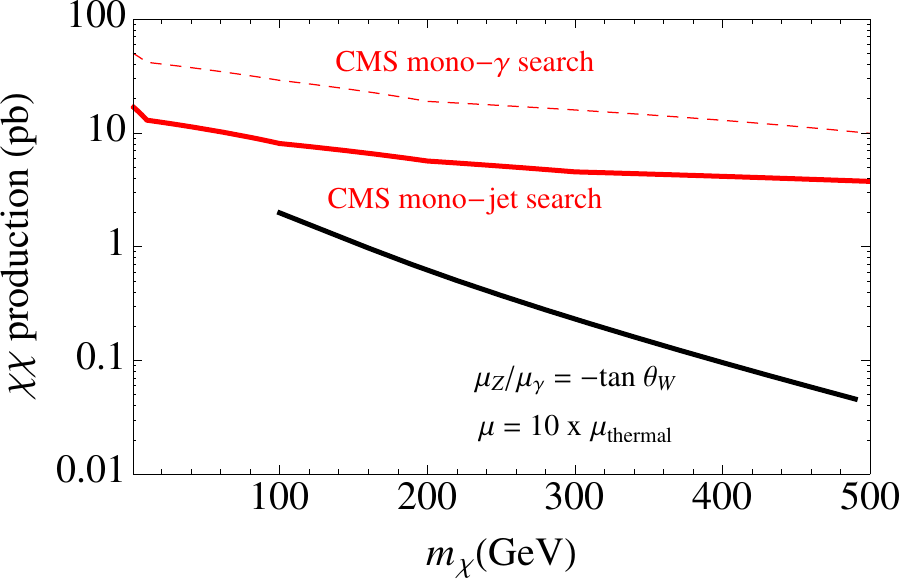}
\end{center}
\caption{A plot of the collider constraints on the production cross-section of a WIMP pair coming from CMS monojet search~\cite{Malik:2012sa} (solid-red) and CMS mono-photon search~\cite{Chatrchyan:2012te} (dashed-red). The constraints from CDF monojet search~\cite{Aaltonen:2012jb} are only slightly weaker compared with the CMS results. In solid black we plot the expected production cross-section in the case of MiDM with a small mass splitting and with a dipole strength ten times larger than $\muther$.}
\label{fig:collider_constraints}
\end{figure}

Limits on RayDM from colliders come primarily from either mono-jet or mono-photon searches. In the case of MiDM with a small splitting the production in colliders is observable only through the emission of a gluon or photon from the initial state partons. Thus the most constraining limits on this scenario come from mono-jet searches. The most recent search from CMS~\cite{Malik:2012sa} place a limit of a few pb in the range $\mX \lesssim 10^{3}\GeV$ (see Fig.~\ref{fig:collider_constraints}). This is not quite sufficient to exclude the interesting production cross-sections in the MiDM scenario (see Fig.~\ref{fig:MiDM_production}), but it comes close. Consulting Tbl.~\ref{tbl:productionVSannihilation} we see for example that in the case of hypercharge dominated  interactions and WIMP mass of $\mX \sim 100\GeV$ monojet searches at the LHC can begin probing annihilation rates into di-photons of the order of $10^{-26}-10^{-27}~{\rm cm^3/s}$.  

On the other hand, when the splitting is large ($\Delta M \gtrsim 100\GeV$) the photon emitted from the excited state's decay is hard enough to be searched for directly. In that case the relevant limits are the limits on $\sigma(pp\rightarrow \slashed{E}  \gamma)$ coming from mono-photon searches. These are much more constraining and are at the level of $ \sigma(pp\rightarrow \chi\chi\gamma) \lesssim 14~{\rm fb}$ for a WIMP mass below the TeV range~\cite{Chatrchyan:2012te}. As can be seen from Fig.~\ref{fig:MiDM_production} this bound all but excludes MiDM with large splittings. In the case of heavy scalar production in association with a photon one can again compare the production rate directly to the limits on $\sigma(pp\rightarrow \slashed{E}  \gamma)$ since the resulting photon is hard. Since the production cross-section in the case of scalars is generically lower (see Fig.~\ref{fig:phi_production}) the current constraints are not quite strong enough to exclude this scenario, but they are now probing the most interesting parts of parameter space.

\section{Conclusions}
\label{sec:conclusions}
The effective theory describing the interactions of a Majorana WIMP with photons is of critical importance, given that our best indirect detection searches come through monoenergetic $\gamma$-ray lines, and direct detection is clearly sensitive to  scattering through a photon exchange. Interestingly, this effective theory is quite restricted: in the presence of a nearby excited state, there is the possibility of an interaction with electromagnetism via a dipole transition to the excited state (or Magnetic Inelastic Dark Matter or MiDM); in the absence of a nearby state, the leading operator comes in the form $\chi \chi W_{\mu\nu} W^{\mu\nu}$ or $\chi \chi B_{\mu\nu} B^{\mu\nu}$ or its the pseudoscalar and CP violating equivalents. These two scenarios  have related, but distinct phenomenology.

Remarkably, in the case of MiDM both the size of the signal in direct detection and $\gamma \gamma$+$\gamma Z$ signatures are independent of the size of the dipole, with the relic abundance suppression precisely canceling out against the enhanced scattering and annihilation cross sections. This offers a surprising concordance whereby the annihilation rates into $\gamma\gamma$ is in the range to explain the tentative excess in gamma rays at around $130\GeV$ and possibly explain the DAMA annual modulation. MiDM predicts a secondary line at around $114\GeV$ from $\gamma\ZZ$ with a relative rate of about $1:3$ compared with the $\gamma\gamma$ line at 130. Fermi should be able to test the $\gamma$ ray signature and, for small mass splitting $\mXp-\mX \approx 100\keV$, the MiDM scenario also predicts collision rates with nuclei that can now be tested at direct detection experiments. The production rates in colliders are below the current sensitivity of the LHC for thermal cross sections, but can exclude some regions of parameter space where this particle constitutes only a fraction of the total dark matter. The case of the thermal WIMP constituting all of the dark matter should be observable in the near future.  We showed that given the scaling of the different quantities involved, the concordance is in fact independent of the dipole strength and is maintained even with an increased dipole strength where MiDM forms only a fraction of the total DM. Model-dependent corrections outside of the effective theory can change this result, however. Moreover, constraints from colliders place an ultimate limit on such an increase in the dipole to be no more than $\mathcal{O}(10)$. 

In the case of RayDM it is also possible to simultaneously achieve the right relic abundance as well as rates in the range now explored by gamma ray observations. But, in contrast with MiDM, it favors stronger coupling to the $\SUWeak$ vector-bosons than to hypercharge. If RayDm is describing only the interactions with photons, however, and freezes out through some other channel, coupling to hypercharge alone gives a good description of the data. RayDM predicts a ratio of $\gamma\ZZ$ to $\gamma\gamma$ of $1:5$ when coupling to hypercharge dominates, or about $5:2$ in the more likely case of dominant coupling to $\SUWeak$. Unfortunately, the direct detection prospects in this case are gloomy as the collision rates with nuclei due to two photon exchange are much too small. In contrast, this scenario offers interesting phenomenology in colliders including mono-photon, mono-Z, and mono-W signatures with rates that can now be probed at the LHC. 

There are a few interesting variations on the scenarios we have discussed. 
A particularly natural scenario is MiDM+RayDM, where the $\gamma \gamma$ signal is naturally boosted in an MiDM model by the presence of an additional hypercharge Rayleigh operator. Such an operator is generally present and would be expected to often dominate the $\gamma \gamma$ signal from these models. 

An alternative possibility is that some amount of hypercharge-dominated RayDM is just a {\em subdominant} component of the dark matter. Since the density scales as $\rho\sim \vev{\sigma v}_{ann}^{-1}$, the overall rate scales as $\rho^2 \vev{\sigma v}_{ann} \sim \vev{\sigma v}_{ann}^{-1}$. Thus, rather than having all dark matter annihilate to $\gamma \gamma$ with a cross section $\vev{\sigma v}_{ann} \sim 3 \times 10^{-27} {\rm cm^3 s^{-1}}$, we could have a cross section $\sim 10 \times \vev{\sigma v}_{thermal}$ and yield the claimed $\gamma \gamma$ signal from a subdominant component of dark matter.

While a number of opportunities exist to distinguish the MiDM scenario from a RayDM scenario, there is another important difference: in RayDM, in particular when the dominant operator is $\chi \chi W_{\mu\nu}W^{\mu\nu}$, there is a sizable hadronic annihilation channel (via $W$'s and $Z$'s) compared to  $\gamma \gamma$. In contrast, for MiDM, the $f \bar f$ channel is not present in the late universe as it is only present for $\chi^* \chi$ annihilations rather than $\chi \chi$. Limits on the continuum photon emissions such as those from dwarf galaxies \cite{GeringerSameth:2011iw,Ackermann:2011wa} or the galactic center \cite{Buchmuller:2012rc,Cohen:2012me,Cholis:2012fb} could potentially distinguish these scenarios.

Ultimately, if a Majorana dark matter interacts significantly with light, there are a number of conclusions that can be drawn right away. While direct detection signals require a nearby state, collider signatures do not. The era of data - approaching dark matter with direct, indirect and collider experiments, may be on the verge of revealing its nature.

\begin{acknowledgments}
We would like to thank Maxim Pospelov for very useful discussions. NW is supported by NSF grant \#0947827. IY is supported in part by funds from the Natural Sciences and Engineering Research Council (NSERC) of Canada. Research at the Perimeter Institute is supported in part by the Government of Canada through Industry Canada, and by the Province of Ontario through the Ministry of Research and Information (MRI). 
\end{acknowledgments}

\onecolumngrid
\appendix


\renewcommand{\theequation}{A-\arabic{equation}}
\setcounter{equation}{0}

\section{Elastic Scattering due to two photons exchange}
\label{app:eval_loop_int}

In this appendix we derive the amplitude for the elastic scattering of the WIMP on the nucleus due to two photons exchange, Eq.~(\ref{eqn:ESAmp}). We begin by noting that there are several separate scales in the problem:  $\LamR$, the high scale associated with the Rayleigh operator;  $\mN$, and $\mX$, the masses of the nucleus and the WIMP, respectively;  ${\bf q}^2$,  the momentum exchange, approximately $100\MeV$;  $\RN$,  the nuclear coherence size approximately $100\MeV$; $q^0$, $E_R$, $\mu v^2$ - the kinetic energies involved, approximately $10\keV$. By itself this diagram is logarithmically divergent, but inclusion of the charge form-factor provides a natural cut-off at a momentum scale around $\RN^{-1}$. Therefore, the momenta running in the loop are non-relativistic and we evaluate this diagram using known techniques from heavy quark effective theory~\cite{Grozin:2000cm}. In appendix~\ref{app:2ndBorn} we show that using this technique one can recover the results obtained in ref.~\cite{Batell:2009vb} where the second order Born cross-section was used to calculate the elastic channel of usual iDM~\cite{TuckerSmith:2001hy}. We begin by writing the momentum of the intermediate nucleus in the usual velocity expansion, 
\be
P = \mN v + \tilde{P},
\ee
where $v = (1,\vec{v})$ is the 4-velocity of the nucleus. The propagator for the nucleus can be approximated as,
\be
\label{eqn:HQETNucleusProp}
\frac{\slashed{P}+\mN}{P^2-\mN^2} \approx  \frac{1}{\tilde{P}\cdot v+i0}\frac{1+\slashed{v}}{2},
\ee
where $(1+\slashed{v})/2$ is the projector onto the two large components of the 4-spinor\footnote{In the diagram above, $P = p+l+q/2$ and since $\tilde{P}\sim l\sim q$ we can write,
\be
\frac{\slashed{p}+\slashed{l}+\slashed{q}/2+\mN}{(p+l+q/2)^2-\mN^2} \approx  \frac{1}{(p+l+q/2)\cdot v+i0}\frac{1+\slashed{v}}{2}
\ee
}. The projector causes the QED vertex of the nucleus to simplify to $i Z e v^{\mu}$ instead of the usual $i Z e\gamma^{\mu}$. This embodies the fact that in the non-relativistic limit, the polarization of charged particles does not change under the exchange of a photon. Given the above, the amplitude associated with this diagram at leading order in the velocity is given by
\be
i\mathcal{M} = \Big( \bar{u}(k') ~u(k)\Big)~ I_{\mu\nu}(q^2)~ \Big(\bar{u}(p') ~\Gamma^{\mu\nu}(v) ~ u(p)\Big),
\ee
with
\be
\Gamma^{\mu\nu}(v) = \left(v^{\mu_2}\left( \frac{1+\slashed{v}}{2}\right) v^{\nu_2}\right), 
\ee
and
\be
\nonumber
I_{\mu_2\nu_2}(q^2) &=& \frac{e^2Z^2g_{\gamma\gamma}}{\LamR^3}\int \frac{d^4l}{(2\pi)^4} \left(\frac{-g_{\mu_1\mu_2}}{(l+q/2)^2} \right)\left(\frac{-g_{\nu_1\nu_2}}{(-l+q/2)^2} \right)\\ \nonumber &\times& \left(\left( -l^2+q^2/4\right)g^{\mu_1\nu_1}-(l+q/2)^{\mu_1}(-l+q/2)^{\nu_1}\right) \\  &\times& \frac{1}{(\tilde{p}+l+q/2)\cdot v + i 0}.
\label{eqn:Imunu}
\ee
Here $I_{\mu_2\nu_2}(q^2)$ has dimensions of inverse mass square. Since we are interested only in the leading order in the velocity only the time-like indices are important and we obtain
\be
\label{eqn:Imunuv0}
I_{{_0}{_0}}(q^2) = \frac{e^2Z^2g_{\gamma\gamma}}{\LamR^3} \int \frac{d^4l}{(2\pi)^4}\frac{{\bf l}^2 - {\bf q}^2/4}{D}~\times~ {\rm F}\left(\left| {\bf l}+\frac{\bf q}{2}\right|\right){\rm F}(\left(\left| -{\bf l}+\frac{\bf q}{2}\right|\right).
\ee
Here, bold face letters denote 3-vectors, and we included the charge form-factor ${\rm F}(|{\bf q}|)$ by hand. The denominator is 
\be
\label{eqn:integral_denominator}
D = (l+q/2)^2(-l+q/2)^2 (\tilde{p}^0+l^0+q^0/2 + i0).
\ee
In order to allow for exact evaluation of this integral we choose to work with the Helm form-factor, which is a function of the momentum exchange ${\bf q}$
\be
F(|{\bf q}|) = e^{-{\bf q}^2/Q_0^2}.
\ee
Here $Q_0 = \sqrt{6} \RN^{-1}$, and the nuclear radius is (see for example the excellent review by Salati~\cite{Salati:2007zz})
\be
\RN =  {\rm fm} \times \left( 0.3 + 0.89 A^{1/3}\right).
\ee
Neglecting the overall constant in front, the integral of Eq.~(\ref{eqn:Imunuv0}) can now be written as
\be
I_{{_0}{_0}}(q^2) \propto \int&~& \frac{d^4l}{(2\pi)^4}\frac{{\bf l}^2 - {\bf q}^2/4}{D} \times \exp\left(- \frac{{\bf l}^2 + {\bf q}^2/4}{Q_0^2}  \right).
\ee
The denominator $D$,  given in Eq.~(\ref{eqn:integral_denominator}), is a factor of several separate propagators that can be combined together using the usual Feynman parameter together with an HQET parameter with the dimensions of energy
\be
\frac{1}{D} = \int_0^1 dx \int_0^\infty d\mathcal{E} \frac{2}{\left(\mathcal{E}\left(\tilde{p}_2^0+l^0+q^0/2\right) + l^2 +q^2/4 + (1-2x) l\cdot q + i\epsilon\right)^3}.
\ee 
Shifting the momentum variable $l_\mu \rightarrow l_\mu+\tfrac{1}{2}\left((1-2x)q_\mu + \mathcal{E} g_{0\mu} \right)$ we can write the integral as
\be
\nonumber
\int dx d\mathcal{E} \frac{d^3{\bf l}}{(2\pi)^3} \int \frac{dl^0}{2\pi} \frac{N}{\left((l^{0})^2 - \Delta + i\epsilon\right)^3} \exp\left(- \frac{{\bf l}^2 -(1-2x){\bf l}\cdot {\bf q} + (2-4x+4x^2){\bf q}^2/4}{Q_0^2}  \right),
\ee
where
\be
N &=& 2\left( {\bf l}^2 - (1-2x) {\bf l}\cdot {\bf q} + x(x-1) {\bf q}^2 \right), \\
\Delta &=& {\bf l}^2 +  \frac{\mathcal{E}^2}{4} + x(x-1) q^2 - \mathcal{E}\left(\tilde{p}_2^0 + xq^0 \right).
\ee
The mixing term in $\Delta$ can be neglected as it is always much smaller than the other two terms. Either way, the integral over $l^0$ can be done exactly and yields
\be
\int_{-\infty}^{\infty} \frac{dl^0}{2\pi} \frac{1}{\left((l^{0})^2 - \Delta + i\epsilon\right)^3} = -\frac{3i}{16\Delta^{5/2}}.
\ee
The dependence of the form-factor on ${\bf l}\cdot {\bf q} = |{\bf l}||{\bf q}|\cos\theta$ causes the integration over $d^3{\bf l}$ to be slightly more complicated than usual. We proceed by first doing the integral over $\cos\theta$ followed by the integral over the dimensional Feynman parameter $\mathcal{E}$. We are left with two integrals, one over the Feynman parameter $x$ and the other over the radial component of the spatial momentum $|\bf{l}|$. Defining the dimensionless variables $\tilde{l} = \left|{\bf  l}\right|/Q_0$ and $\tilde{q} = \left|{\bf  q}\right|/Q_0$ we arrive at the result quoted in the text in Eq.~(\ref{eqn:ESAmp})
\be
i\mathcal{M} =  \frac{i \alpha Z^2g_{\gamma\gamma}}{4}\frac{Q_0}{\LamR^3}~\mathcal{F}\left( \frac{\left|{\bf q}^2\right|}{Q_0^2}\right)~ \bar{u}(k') u(k)~ \bar{u}(p') u(p),
\ee
where,
\be
\mathcal{F}\left( \tilde{q} \right) = \frac{4}{\pi} \int_0^{1} dx \int_0^\infty &d\tilde{l}& \frac{\tilde{l}^2 }{(\tilde{l}^2 + (1-x)x \tilde{q}^2)^2} \times \exp\left(-\tilde{l}^2 - \tilde{q}^2(\tfrac{1}{2}-x+x^2) \right) \\ \nonumber &\times&\left(\cosh\left((1-2x)\tilde{l}\tilde{q} \right) -  \frac{\tilde{l}^2 - (1-x)x\tilde{q}^2+1}{(1-2x)\tilde{l}\tilde{q}} ~\sin\left((1-2x)\tilde{l}\tilde{q} \right)\right).
\ee

\renewcommand{\theequation}{B-\arabic{equation}}
\setcounter{equation}{0}

\section{Derivation of the elastic channel of iDM using HQET}
\label{app:2ndBorn}

\begin{figure}
\begin{center}
\includegraphics[width=0.35 \textwidth]{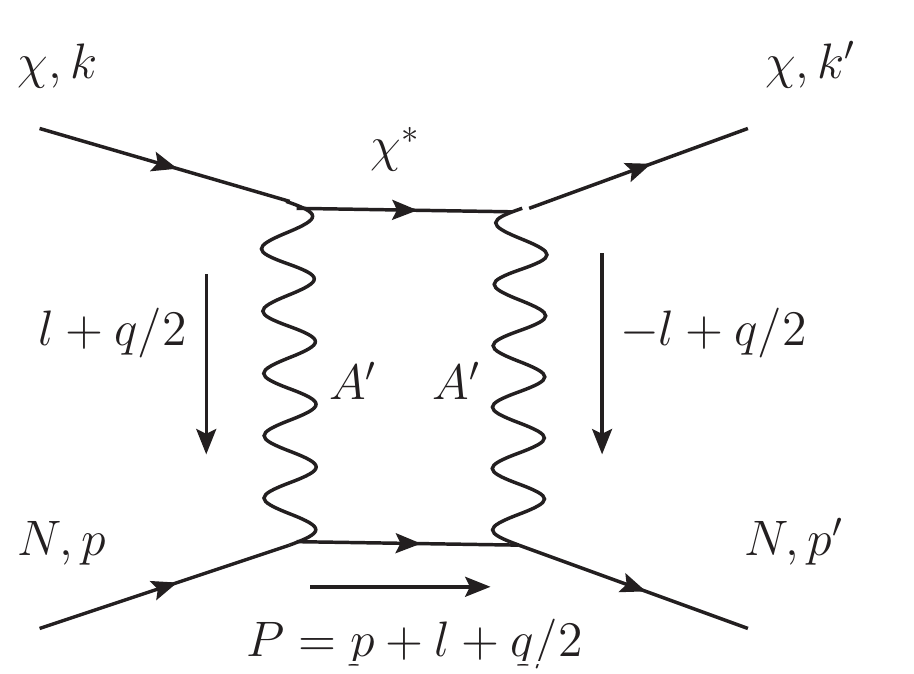}
\end{center}
\caption{The second Born amplitude for the process $\chi N\rightarrow\chi N$ through an intermediate excited state $\chi^*$. A second diagram with the $A'$ lines crossed contributes the same as this one. }
\label{fig:ES_second_order}
\end{figure}

The inelastic Dark Matter scenario of Ref.~\cite{TuckerSmith:2001hy} involves a dark matter state $\chi$ that interacts with normal matter only through a transition involving an excited state $\chi^*$ separated in mass by $\DeltaM$. Ref.~\cite{Batell:2009vb} considered the particular case where the interaction with the SM is through a new massive U'(1) vector-boson that kinetically mixes with hypercharge, a Holdom boson~\cite{Holdom:1985ag}. They computed for the first time the contribution to the elastic channel from the second-order diagram shown in Fig.~\ref{fig:ES_second_order} in the case where the mass splitting is much larger than the kinetic energy available, $\DeltaM \gtrsim \MeV$. In what follows, we reproduce this result using the HQET methods discussed in the text and appendix~\ref{app:eval_loop_int}. For simplicity we ignore the charge form-factor. 

Writing the four-momentum of the WIMP as $k = \mX u + \tilde{k}$ with $u = (1,\vec{u})$ being the 4-velocity, we can express the momentum of the excited state as,
\be
K = k - l -q/2 = \mX u + \tilde{k} -l -q/2.
\ee
Since the mass splitting, $\DeltaM$ is greater than the kinetic energy, the fermionic propagator of the excited state can be written as,
\be
\frac{\slashed{K}+\mXp}{K^2-\mXp^2} \approx  \frac{1}{\DeltaM}\frac{1+\slashed{u}}{2}.
\ee
The propagator for the nucleus is as given in Eq.~(\ref{eqn:HQETNucleusProp}). Replacing $\gamma^\mu$ by four-velocities $v^\mu$ in the vertices and considering the leading order term in the velocity expansion the amplitude is given by
\be
i\mathcal{M} = \left( \bar{u}(k')~u(k)\right)~ I(q^2)~ \left(\bar{u}~ u(p)\right),
\ee
with,
\be
I(q^2) = 2 \times \frac{16\pi^2 \kappa^2 Z^2 \alpha \alpha'}{\DeltaM} \int \frac{d^4l}{(2\pi)^4} \frac{1}{D}.
\ee
Here the factor of $2$ arises from a second diagram where the $A'$ lines cross and which contributes equally at this order. Here $\kappa$ is the kinetic mixing parameter, $\alpha' = e^{'2}/4\pi$ is the U'(1) charge. $D$ is the same denominator as previously considered and presented in Eq.~(\ref{eqn:integral_denominator}) except that the massless photon propagators are replaced by a massive propagator $(q^2+i\epsilon)^{-1} \rightarrow (q^2-m_{A'}^2+i\epsilon)^{-1}$. Following the same steps as in appendix~\ref{app:eval_loop_int}, we introduce the Feynman parameters $x$ and $\mathcal{E}$ to combine the denominator, shift the loop momentum $l_\mu \rightarrow l_\mu+\tfrac{1}{2}\left((1-2x)q_\mu + \mathcal{E} g_{0\mu} \right)$, and then integrate over $dl^0$, 
\be
I(q^2) = -\frac{3i}{4}\left(\frac{16\pi^2 \kappa^2 Z^2 \alpha \alpha'}{\DeltaM} \right) \int_0^1 dx \int_0^\infty d\mathcal{E} \int\frac{d^3{\bf l}}{(2\pi)^3}\frac{1}{\left({\bf l}^2 + \Delta'\right)^{5/2}}.
\ee
Here $\Delta' = \mathcal{E}^2/4 + x(x-1)q^2 + m_{A'}^2 - \mathcal{E}(\tilde{p}^0 + xq^0)$. Neglecting the term linear in $\mathcal{E}$ as before, all the integrals can easily be done  to yield 
\be
I(|{\bf q}|) = -\frac{i}{4\pi}\left(\frac{16\pi^2 \kappa^2 Z^2 \alpha \alpha'}{\DeltaM} \right) \frac{{\rm Arctan}\left(|{\bf q}|/2m_{A'}\right)}{|{\bf q}|} \quad \overset{m_{A'}\rightarrow 0}{\longrightarrow} \quad \frac{2 \pi^2 Z^2 \alpha \alpha' \kappa^2}{|{\bf q}| \DeltaM}.
\ee
The differential cross-section in the CM frame is simply
\be
\frac{d\sigma}{d\Omega} = \frac{\mu^2}{4\pi^2}I^2(|{\bf q}|) = \frac{4 \alpha^2\alpha{'2}Z^4\kappa^4 \mu^2}{{\bf q}^2\DeltaM} {\rm Arctan}\left(\frac{|{\bf q}|}{2m_{A'}}\right)^2\quad \overset{m_{A'}\rightarrow 0}{\longrightarrow} \quad \frac{\pi^2 \alpha\alpha'Z^2\kappa^2 \mu^2}{{\bf q}^2\DeltaM},
\ee
which reproduces the result obtained in ref.~\cite{Batell:2009vb} in the appropriate limit. 


\renewcommand{\theequation}{C-\arabic{equation}}
\setcounter{equation}{0}

\section{Differential Cross-Section Formulas for MiDM and RayDM Production at Colliders}
\label{app:collider_production}

In this appendix we give the formulas for the production cross-section and distributions for the process $f\bar{f}\rightarrow \bar{\chi}\chi V$ where $V = \gamma,\ZZ, \Wpm$ through the Rayleigh operators of Eq.~(\ref{eqn:RDMLagrangian}) as well as related processes in UV completions of RayDM. A convenient way of presenting the results is obtained by treating the $\bar{\chi}\chi$ system as having four-momentum $p_4$ and mass $p_4^2$. The $2\rightarrow 3$ phase-space factor can then be written as
\be
{\rm dPS}_3\left(p_{f}p_{\bar{f}}\rightarrow p_{V} + p_{\bar{\chi}}+ p_{\chi}\right)  = {\rm dPS}_2\left(p_{f}p_{\bar{f}}\rightarrow p_{V} + p_4\right) \times \frac{dp_4^2}{2\pi} \times  {\rm dPS}_2\left(p_4\rightarrow  p_{\bar{\chi}}+ p_{\chi}\right).
\ee
Averaging (summing) over initial (final) polarization for the process shown in Fig.~\ref{fig:XXprodRDM}, and neglecting the incoming particles' masses, the matrix element squared is
\be
\label{eqn:productionM2}
\frac{1}{4}\sum_{\rm pol} \left| \mathcal{M}\right|^2 &=& \frac{(g_v^2 + g_a^2) g_{VV}^2 ~s~p_4^2}{\LamR^6} \frac{1}{2 \left(s-m_V^2\right)^2} ~\times \\ \nonumber &~& \Big[ ~ a_S^2 \left(1 -4\mX^2/p_4^2\right)\left(m_V^4 + (s-p_4^2)^2 + 2 m_V^2 (s - t) + 2 (s-p_4^2) t + 2 t^2 \right)  \\ \nonumber &\quad& + a_A^2\left( m_V^4 + (s-p_4^2)^2 + 2 (s-p_4^2) t + 2 t^2 - 2 m_V^2 (s + t) \right) \Big],
\ee
where $g_{v,a}$ are the vector and axial couplings of $V$ to $\bar{f}f$, $g_{VV}$ are defined in Eq.~(\ref{eqn:gVV}), and the Mandelstam variables $s=\left(p_{f}+p_{\bar{f}}\right)^2$, $t=\left(p_{f}-p_{V} \right)^2$, and $u=\left(p_{f}-p_4 \right)^2$ are defined as usual~\cite{Peskin:1995ev}. Here we kept explicit the separate contributions from the scalar $\bar{\chi}\chi FF$ (axial $\bar{\chi}\gamma_5\chi F\tilde{F}$) piece by preceding it with $a_S$ ($a_A$). This separation will prove useful below when discussing the corresponding formulas in the UV completion of RayDM with heavy scalars. The integral over  ${\rm dPS}_2\left(p_4\rightarrow  p_{\bar{\chi}}+ p_{\chi}\right)$ is straightforward and can be done in its entirety since the squared amplitude in Eq.~(\ref{eqn:productionM2}) contains no dependence on the $\bar{\chi}\chi$ system's angular distribution. The integral over the azimuthal angle of $ {\rm dPS}_2\left(p_{f}p_{\bar{f}}\rightarrow p_{V} + p_4\right)$ can also be done and the differential cross-section is then given by
\be
\nonumber
\frac{d^2\sigma(f\bar{f}\rightarrow \bar{\chi}\chi V)}{d\cos\theta ~dp_4^2} &=&  \frac{(g_v^2 + g_a^2) g_{VV}^2~p_4^2}{2048\pi^3\LamR^6 \left(s-m_V^2\right)^2} ~\left(1+\cos^2\theta \right)~ \times ~ \sqrt{1-\tfrac{4\mX^2}{p_4^2}}  ~\sqrt{\lambda\left(1,\frac{m_V^2}{s},\frac{p_4^2}{ s}\right)} \\ \nonumber &~& \Bigg[ ~ a_S^2  \left(1 -4\mX^2/p_4^2\right)  \left( m_V^4 + (p_4^2 - s)^2 + 
 2 m_V^2 \left(\left(\frac{3 - \cos^2\theta}{1 + \cos^2\theta}\right)~s-p_4^2\right)\right) \\ &\quad& + a_A^2\left( m_V^4 + (s-p_4^2 )^2 - 2 m_V^2 (p_4^2 + s)\right) \Bigg],
\ee
where $\theta$ is the angle between the incoming fermion $f$ and the vector $V$ in the center-of-mass frame, and $\lambda(x,y,z) = x^2+y^2+z^2 - 2x y - 2y z - 2zx$ is the usual kinematic function. The angular region is $-1 \le \cos\theta < 1$ and the mass of the $\bar{\chi}\chi$ system is in the range $4\mX^2 < p_4^2 < \left(\sqrt{s}-m_V^2\right)^2$. The photon case is particularly simple and yields 
\be
\frac{d^2\sigma(f\bar{f}\rightarrow \bar{\chi}\chi\gamma)}{d\cos\theta dp_4^2} &=&\frac{\alpha g_{\gamma\gamma}^2}{512\pi^2} ~\frac{p_4^2\left(1+\cos^2\theta \right)}{\LamR^6} \sqrt{1-\tfrac{4\mX^2}{p_4^2}}\left(a_A^2 + a_S^2\left(1 -4\mX^2/p_4^2\right) \right)\left( 1- \frac{p_4^2}{s}\right)^{3/2}.
\ee
This can be integrated exactly to yield the total cross-section that is quoted in Eq.~(\ref{eqn:ffbartotxs}) for the case where $\sqrt{s} \gg \mX$. What is often of more interest is the transverse momentum distribution of the vector boson.  The transverse momentum in the centre-of-mass frame is given by $p_T = p_V \sin\theta$ and the momentum of the vector in the center-of-mass frame is $p_V =  \frac{\sqrt{s}}{2}  \lambda^{1/2}\left(1,\frac{m_V^2}{s},\frac{p_4^2}{ s}\right)$. This can be used to obtain the differential distribution 
\be
\frac{d\sigma}{dp_T} = \int dp_4^2 ~d\cos\theta ~ \frac{d^2\sigma}{d\cos\theta dp_4^2} ~\delta\left(p_T(\cos\theta,p_4^2)-p_T \right).
\ee
Needless to say, for hadronic colliders such as the LHC and the Tevatron these expressions have to be convolved against the appropriate parton distribution functions. We have verified the formulas above against the Madgraph 4 package~\cite{Alwall:2007st}.

Similarly, in the case of heavy scalars, by putting the $\bar{\chi}\chi$ momentum on-shell $p_4^2 =m_{\phiS}^2$ we obtain the matrix-element squared for the process $f\bar{f}\rightarrow \gamma \phiS$ through $\gamma/\ZZ$
\be
\label{eqn:phiProdM2}
\frac{1}{4}\sum_{\rm pol} \left| \mathcal{M}\right|^2 &=& \frac{\pi \alpha~ q_f^2}{\Lambda_{\rm UV}^2} ~\frac{t^2+u^2}{s} \times \Big(g_{\gamma\gamma}^2 + 2 g_{\gamma\gamma} g_{\gamma\ZZ} v_f \xi(s) + g_{\gamma\ZZ}^2 (v_f^2+a_f)^2 \xi^2(s)\Big).
\ee
Here $s,t$, and $u$ are the usual Mandelstam variables. Use of the relation
\be
\frac{d\sigma}{dp_T} = \left(\frac{p_T}{p}\frac{1}{\sqrt{p^2 - p_T^2}} \right) \frac{d\sigma}{d\cos\theta},
\ee
leads to the differential cross-section given by Eq.~(\ref{eqn:ffTOphigammaDiff}). Here $p=\frac{\sqrt{s}}{2} \left(1 - \frac{m_{\phiS}^2}{s}\right)$ is the momentum of the photon in the center of mass frame. Similar expressions hold for the other distributions involving $\ZZ$ and $\Wpm$. 

Finally, in the case of MiDM, the production cross-section is given by
\be
\label{eqn:MiDMprodXS}
&\frac{d\sigma\left( f\bar{f}\rightarrow \bar{\chi}\chi \right)}{d\cos\theta} & = \frac{\alpha~ q_f^2~\mug^2}{8} \left(1 + 2 v_f \frac{\muZ}{\mug} \xi(s) + (v_f^2+a_f^2) \frac{\muZ^2}{\mug^2} \xi^2(s)\right) \\ \nonumber &\times& \sqrt{\lambda\left(1,\frac{\mX^2}{s},\frac{\mXp^2}{ s}\right)}  \times \left(1-\frac{\Delta M^2}{s}\right) \left(\sin^2\theta + \frac{\left(\mX+\mXp\right)^2}{s}\left(1+\cos^2\theta\right) \right), 
\ee
where $\theta$ is the scattering angle in the centre of mass frame, and $v_{\gamma,\ZZ}$ ($a_{\gamma,\ZZ}$ ) is the vector (axial) coupling of the corresponding vector-boson to the incoming fermions.

\bibliography{RayDMbib}

\end{document}